\documentclass{aastex}
\usepackage{emulateapj5}    
\usepackage{graphicx}
\usepackage{timesfonts}     
\newcommand{\be}{\begin{equation}}
\newcommand{\ee}{\end{equation}}
\newcommand{\bea}{\begin{eqnarray}}
\newcommand{\eea}{\end{eqnarray}}

\begin{document}
\title{Statistics of Galactic Synchrotron and Dust Foregrounds:
  Spectra, PDFs and Higher-Order Moments}
\shorttitle{Synchrotron and Polarized Dust Foregrounds}
\shortauthors{Cho \& Lazarian}

\author{Jungyeon Cho\altaffilmark{1,2} 
\and
A. Lazarian\altaffilmark{2}
}
\altaffiltext{1}{Dept. of Astronomy and Space Science, 
    Chungnam National Univ., Daejeon, Korea; cho@canopus.cnu.ac.kr}
\altaffiltext{2}{Dept. of Astronomy, Univ. of Wisconsin, Madison, 
    WI53706, USA; cho@..., lazarian@astro.wisc.edu}

\begin{abstract}

We present statistical analysis of diffuse Galactic synchrotron emission
and polarized thermal emission from dust.
Both Galactic synchrotron emission and
polarized thermal emission from dust reflect statistics of
magnetic field fluctuations and, therefore, Galactic turbulence.
We mainly focus on the relation between observed angular spectra and
underlying turbulence statistics.
Our major findings are as follows.
First, we find that magnetohydrodynamic (MHD) turbulence in the Galaxy
can indeed explain 
diffuse synchrotron emission from high galactic latitude.
Our model calculation suggests that either a one-component
extended halo model or a two-component model, 
an extended halo component (scale height $\gtrsim 1kpc$) plus 
a local component, can explain the observed angular spectrum of
the synchrotron emission.
However, discrete sources seem to dominate the spectrum for
regions near the Galactic plane.
Second, we study how star-light polarization is related with
polarized emission from thermal dust.
We also discuss the expected angular spectrum of polarized emission
{}from thermal dust.
Our model calculations suggest that $C_l\propto l^{-11/3}$
for $l\gtrsim 1000$ and a shallower spectrum for $l\lesssim 1000$.

\end{abstract}

\section{Introduction}
Diffuse Galactic synchrotron emission is an important foreground source of 
the cosmic microwave background (CMB) signals.
Therefore proper understanding of Galactic synchrotron emission
is essential for CMB studies.
In this paper, we present statistical analysis of a synchrotron foreground
emission map. In particular, we 
provide physical interpretation of the angular
 spectrum of
the synchrotron intensity.
In addition, we briefly discuss the properties
of a model dust emission map, another important ingredient of
the CMB foreground.
As measurements of the polarized CMB signals become possible,
 more accurate  removal of Galactic polarized foreground emission is
required.
Synchrotron and dust emissions are known to be the most important
sources of polarized foreground radiation.
Therefore, measurements of angular power spectra of such foregrounds are
of great interest.
In this paper, we also provide estimation of
angular spectrum of polarized 
emission by foreground dust.

The angular spectrum of the synchrotron emission delivers valuable information
on the structure of the Galaxy.
The observed spectra of synchrotron emission 
and synchrotron polarization
(see papers in de Oliveira-Costa \& Tegmark 1999)
reveal a range of power-laws.
Since the Galactic synchrotron emissivity is roughly proportional to
the magnetic energy density \footnote{
    In fact, correct scaling is
    $\epsilon({\bf r}) \propto n(e) |{B}_{\bot}|^{\gamma}$,
    where $\epsilon({\bf r})$ is the synchrotron emissivity,
    $n(e)$ is the cosmic-ray electron number density, and
    $B_{\bot}$ is the strength of the magnetic field component
    perpendicular to the line of sight.
    In our Galaxy, we have $\gamma \sim 2$.
    Therefore, when $n(e)$ is more or less uniform, we have
    $\epsilon({\bf r}) \propto  |{B}_{\bot}|^{2}$.},
angular spectrum of synchrotron emission reflects statistics of
magnetic field fluctuations in the Galaxy 
(see \S\ref{sect:synchem} for further discussions).

The interstellar medium (ISM) is turbulent and Kolmogorov-type spectra
were reported on the scales from several AUs to several kpc
(see Armstrong, Rickett, \& Spangler 1995; Lazarian \& Pogosyan 2000; Lazarian 2009). 
It is believed that magnetic field lines are twisted and bend by
turbulent motions in the Galaxy.
Therefore it is natural to think of the 
turbulence as the origin of the magnetic field fluctuations
and thus the diffuse synchrotron foreground
radiation. Indeed, several earlier studies addressed this issue.
Tegmark et al.~(2000) suggested that the spectra may be
relevant to Kolmogorov turbulence.
Chepurnov (1999) and Cho \& Lazarian (2002; hereinafter CL02) used
different approaches, but both showed that 
the angular spectrum of synchrotron
emission reveals Kolmogorov spectrum ($C_l\propto l^{-11/3}$) for large values 
of multipole $l$.
However, they noted that
the spectrum can be shallower than the Kolmogorov one for 
intermediate values of multipole $l$, due to density stratification
in the halo (Chepurnov 1999)
or the Galactic disk geometry (CL02).
In this paper, we further elucidate the relation between 3-dimensional
turbulence spectrum and observed angular spectrum.
We also investigate how structure of the Galactic halo affects
observed angular spectrum.

Thermal emission from dust is also an important source of
foreground emission.
Dust emission is the most
pronounced emission in far infrared (FIR) wavelengths.
Schlegel, Finkbeiner, \& Davis (1998) combined 100$\mu m$ maps of
IRAS (Infrared Astronomy Satellite) and 
DIRBE (Difuse Infrared Background
Experiment on board the COBE satellite) and removed the
zodiacal foreground and point sources to
construct a full-sky map.
Finkbeiner, Davis, \& Schlegel (1999) extrapolated the 100$\mu m$ emission
map and 100/240$\mu m$ flux ratio maps to sub-millimeter and
microwave wavelengths.
In this paper, we present statistical analysis of a
model dust emission map above.

Thermal radiation from dust becomes polarized when
dust grains are aligned.  
There is ample evidence that dust grains are aligned with respect
to magnetic field.
Therefore, polarization by dust is also related to
the magnetic field fluctuations.
Then how are they related?
Polarized dust emission is difficult to observe directly.
Therefore in this paper we first study the relation between
star-light polarization and turbulence statistics.
Then we describe how we infer the relation between
star-light polarization and polarized thermal emission from dust.

In this paper, we show how 
MHD turbulence is related with the observed
angular spectra of Galactic synchrotron emission  and
starlight polarization.
For this purpose, we use an analytical insight
obtained in Lazarian (1992, 1995ab) and numerical results
obtained in CL02.
We also discuss  how we can estimate polarized microwave dust emission
using star-light polarimetry.
This problem is of great importance in view of recent interest to
the foregrounds to the CMB polarization.
In \S2, we briefly describe the data sets we use in this paper, 
the 408MHz Haslam map and a model
dust emission map at 94GHz.
In \S3, we review a simple model of the angular spectrum of 
synchrotron emission arising from homogeneous MHD turbulence.
In \S4, we present statistical analysis of the Haslam map, which
   is dominated by diffuse Galactic synchrotron emission.
In \S5, we investigate the dust emission map.
In \S6, we consider polarized emission from thermal dust.
In \S7, we discuss how to utilize our results
 to remove Galactic foregrounds.
In \S8, we calculate high-order structure functions of
 the synchrotron and the dust maps and we compare the results with
 those of turbulence.
Finally in \S9, we give summary.

\section{Data Sets}
We use the 408MHz Haslam all-sky map (Haslam et al.~1982)
and a model dust emission map 
that are available on the NASA's LAMBDA website \footnote{
          http://lambda.gsfc.nasa.gov/}.
Both maps were reprocessed for HEALPix (G\'{o}rski et al. 2005)
with nside=512.

The original Haslam data were produced by merging several different data-sets.
``The original data were 
processed in both the Fourier and spatial domains to mitigate baseline 
striping and strong point sources'' (see the website for details). 
The angular resolution of the original Haslam map is $\sim 1^{\circ}$.
Galactic diffuse synchrotron emission is the dominant source of
emission at 408MHz.

The 94 GHz dust map 
is based on fits to data from earlier 100 micron and 100/240 micron maps (Schlegel et al. 1998) and extension to COBE/FIRAS frequencies and identical to the two-component model 8 of 
Finkbeiner et al. (1999).

\section{Spectrum and Structure Function of Diffuse Synchrotron Emission: a model for homogeneous turbulence revisited} \label{sect:spsfco}
Suppose that 3-dimensional MHD turbulence
has a 3D spatial power spectrum of the form $E_{3D}\propto k^{-m}$, 
where $k$ is the wavenumber. Note that in Kolmogorov turbulence
$m=11/3$.
Then what will be the 
2-dimensional angular spectrum, $C_{l}$, of the observed synchrotron total intensity?
We mostly follow discussions in Cho \& Lazarian (CL02).
Although we focus on synchrotron emission here, 
the discussion in this section
can be applicable to any kind of emission
{}from an optically thin medium.

\subsection{MHD turbulence and synchrotron emission} \label{sect:synchem}

For synchrotron radiation, emissivity at a point ${\bf r}$ is given by
   $\epsilon({\bf r}) \propto n(e) |{B}_{\bot}|^{\gamma}$,
where $n(e)$ is the electron number density, $B_{\bot}$ is the component
of magnetic field perpendicular to the line of sight. The index
$\gamma$ is approximately $2$ for radio synchrotron frequencies
(see Smoot 1999).
If electrons
are uniformly distributed over the scales of magnetic field inhomogeneities,
the spectrum of synchrotron
intensity reflects the statistics of magnetic field.
For small amplitude perturbations
($\delta b/B\ll 1$; this is true for scales several times 
smaller than the outer scale of turbulence 
if we interpret $B$ as local mean magnetic field strength and $\delta b$ as random
fluctuating field in the local region), 
if $\delta b$
has a power-law behavior, the synchrotron emissivity will have the 
same power-law behavior (see Getmantsev 1959; Lazarian \& Shutenkov 1990;
Chepurnov 1999).
Therefore, we expect that the angular spectrum of synchrotron
intensity also reflects the spectrum of 3-dimensional MHD turbulence.

When an observer is located inside a turbulent medium, the angular 
correlation function, hence the power spectrum,
 shows two asymptotic behaviors. When the angle is larger than
a critical angle, we can show that 
the angular correlation shows a universal $\theta^{-1}$
scaling. On the contrary, when the angular separation
 is smaller than the critical angle, 
the angular correlation
reflects statistics of turbulence. In this small angle limit,
we can show that
the angular power spectrum is very similar to that of turbulence.
The critical angle is determined by the geometry.
Let the outer scale of turbulence be $L$ and the distance to the farthest
eddies be $d_{max}$.
Then the critical angle is 
\be
  \theta \sim L/d_{max}.
\ee

\subsection{Small-angle limit in homogeneous turbulence} \label{sect:smangle}
When the angle between the lines of sight
is small (i.e.~$\theta < L/d_{max}$), the
angular spectrum $C_l$ has the same slope as the 3-dimensional
energy spectrum of turbulence.
 Lazarian \& Shutenkov (1990) showed that if the 3D spatial spectrum
of a variable follows
a power-law, $E_{3D}(k)\propto k^{-m}$, 
then the 2-dimensional spectrum of the variable projected on the sky also
 follows the same power-law,
\begin{equation}
   C_l \propto   l^{-m}
\end{equation}
in the small $\theta$ limit.
For Kolmogorov turbulence ($E_{3D}\propto k^{-11/3}$),
we expect
\begin{equation}
C_l \propto l^{-11/3}, \mbox{~~~if $\theta<L/d_{max}$.}
 \label{eq_5}
\end{equation}
Note that $l\sim \pi/\theta$.

In some cases, when we have data with incomplete sky coverage,
we need to infer $C_l$ from the observation of 
    the angular correlation function 
\be
  K(\theta)=< T({\bf e}_1) T({\bf e}_2) >,
\ee
where ${\bf e}_1$ and ${\bf e}_2$ are unit vectors along the lines of
sight, $\theta$ is the angle between ${\bf e}_1$ and ${\bf e}_2$, and
the angle brackets denote average taken over the observed region.
As we discuss in Appendix A, when the underlying 3D turbulence spectrum
is $\propto k^{-m}$ (e.g. $m=11/3$ for Kolmogorov turbulence),
the angular correlation function $K(\theta)$ is given by
\be
   K(\theta) ~\propto ~< T^2 >-\mbox{const}~\theta^{m-2}, \mbox{~~~if $\theta < L/d_{max}$.}
\ee
It is sometimes
inconvenient to use the angular correlation function in practice to
study turbulence statistics
because of the constant $< T^2 >$.

A better quantity in small-angle
limit would be the second-order angular structure function:
\bea
   D_2 (\theta) & = & < | T({\bf e}_1) -T({\bf e}_2) |^2 > \\
     & = & 2< T^2 > - 2K(\theta).
\eea
Thus, in homogeneous turbulence with 3D spatial spectrum 
of $E(k)\propto k^{-m}$, we have
\be
   D_2 (\theta) \propto \theta^{m-2}.
\ee
When we measure the slope of the angular structure function, we
can infer the slope of the 3D spatial power spectrum of turbulence.

\begin{figure*}[h!t]
\includegraphics[width=0.40\textwidth]{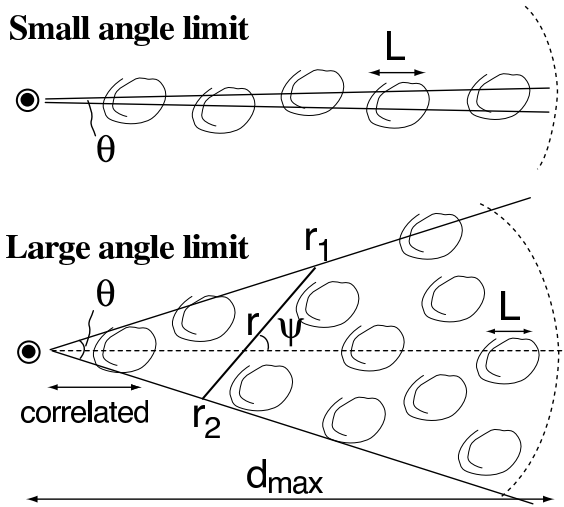}  
\caption{Two limits in homogeneous turbulence.
  {\it Upper plot}:
Small $\theta$ limit ($\theta < L/d_{max}$). 
The fluctuations along the entire length of the lines of sight
are correlated. 
{\it Lower plot}: Large $\theta$ limit ($\theta > L/d_{max}$). 
Only points close to the observer
are correlated. Note the definition of
$r$ and $\psi$. From CL02.
}
\label{fig:0}
\end{figure*}

\subsection{Large-angle limit in homogeneous turbulence}
In this limit, the angular correlation function is 
more useful than the structure function.
{}Following Lazarian \& Shutenkov (1990), we can show that
the angular correlation function for $ \theta > L/d_{max}$ follows
\begin{eqnarray}
    K(\theta) & = & \int \int dr_1 dr_2 ~{\cal K}( |{\bf r}_1-{\bf r}_2| ), \nonumber
\\
              & = & \frac{1}{\sin{\theta}}
                    \int_0^{\infty}dr ~r {\cal K}(r) \int_{\theta/2}^{\pi-\theta/2}d\psi 
\propto \frac{\pi-\theta}{\sin\theta} \sim \frac{const}{\theta},
\end{eqnarray}
where we change variables: $(r_1,r_2)\rightarrow (r,\psi)$, 
which is clear from
Fig.~\ref{fig:0}. We accounted for the Jacobian of which is $r/\sin{\theta}$.
We can understand $1/\theta$ behavior qualitatively as follows.
When the angle is large, points along  of the lines-of-sight near the observer
are still correlated. These points extend from the observer
over the distance $\propto 1/\sin{(\theta/2)}$.

If we assume $L/d_{max}< \theta\ll 1$,
we can get the angular power spectrum $C_l$ 
using Fourier transform:
\begin{eqnarray}
   C_l & \sim &  \int \int  K(\theta) 
                 e^{-i{\bf l}\cdot {\bf \theta}} d\theta_x d\theta_y  
\nonumber \\
       & \sim &  \int d\theta ~\theta J_0(l\theta) K(\theta) \propto   l^{-1},
   \label{eq_7}
\end{eqnarray}
where $\theta=(\theta_x^2+\theta_y^2)^{1/2}$, $J_0$ is the Bessel function, and
we use $K(\theta)\propto \theta^{-1}$.

\subsection{Expectations for homogeneous turbulence}
In summary, for homogeneous Kolmogorov turbulence, 
we expect from equations (\ref{eq_5}) and
(\ref{eq_7}) that
\begin{equation}
 C_l \propto \left\{ \begin{array}{ll} 
                         l^{-11/3}     & \mbox{if $l>l_{cr}$} \\
                         l^{-1}        & \mbox{if $l<l_{cr}$,}
                      \end{array}
              \right.
     \label{eq_1_11_3}
\end{equation}
which means that the power index $\alpha$ of $C_l$ is\footnote{ 
      Note that point sources would result in $\alpha \sim 0$.} 
$-1 \leq \alpha \leq -11/3$.
For small-angle limit, we expect the following scaling for 
the second-order angular structure function:
\be
 D_2(\theta) \propto   \theta^{5/3}     \mbox{~~~if $\theta <L/d_{max}.$} 
     \label{eq_d2}
\ee 
The critical angle $\theta_{cr} \sim L/d_{max}$ 
depends on the size of the large
turbulent eddies and on the length of the line of sight. 
If we assume that turbulence is homogeneous along the lines
of sight and has $L\sim 100\ pc$ corresponding to a typical size of the
supernova remnant, and that $d_{max}\sim 1$~kpc for synchrotron halo (see Smoot 1999), 
we get $\theta_{cr} \sim 6^{\circ}$.

\section{Properties of Diffuse Galactic Synchrotron Emission}
In this section, we analyze the Haslam 408MHz all-sky map, which is
dominated by Galactic diffuse synchrotron emission.
Our main goal is to explain the observed synchrotron angular spectrum
using simple turbulence models.


\subsection{General properties of the 408MHz Haslam map}
\begin{figure*}[h!t]
\includegraphics[width=0.40\textwidth]{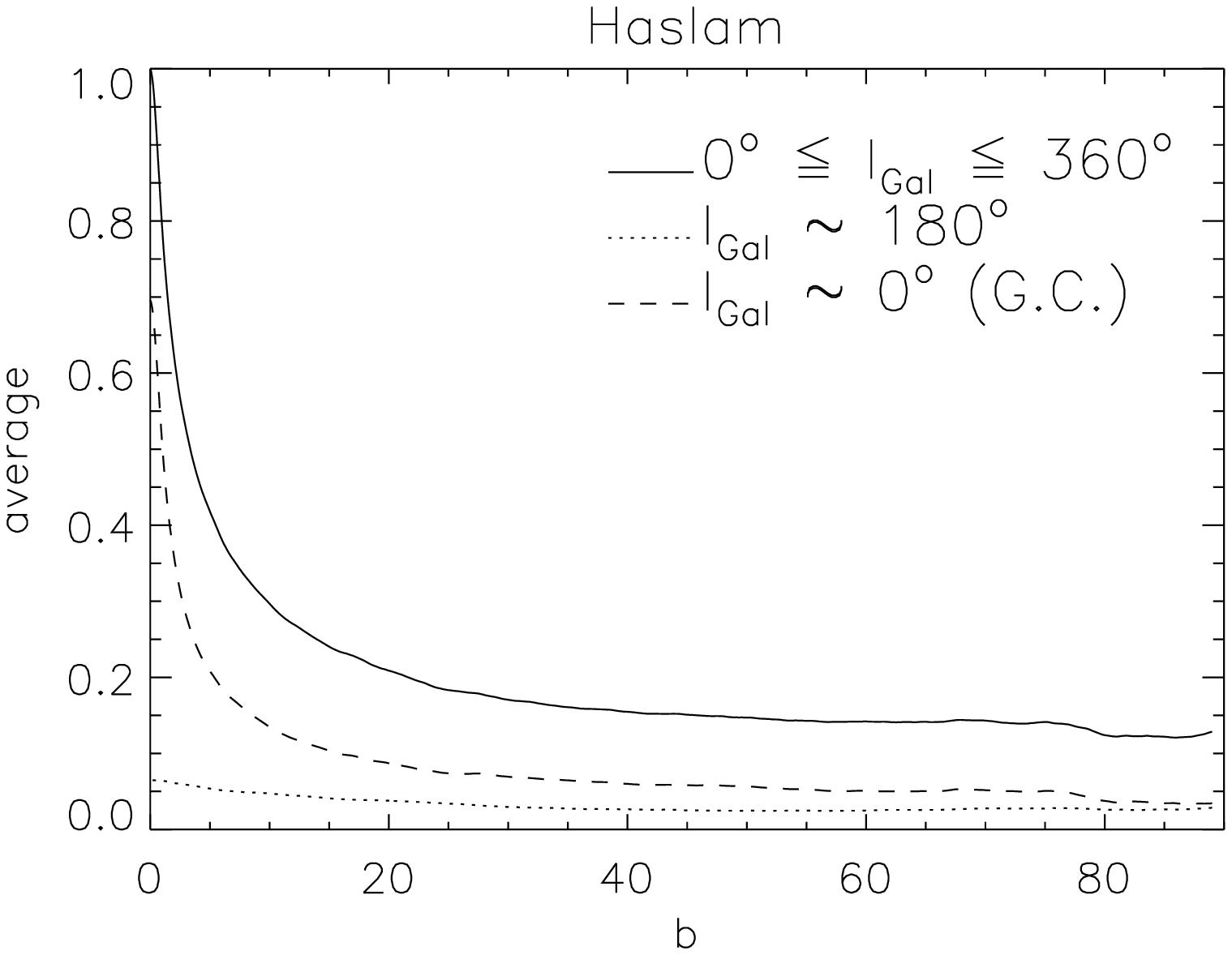}
\includegraphics[width=0.40\textwidth]{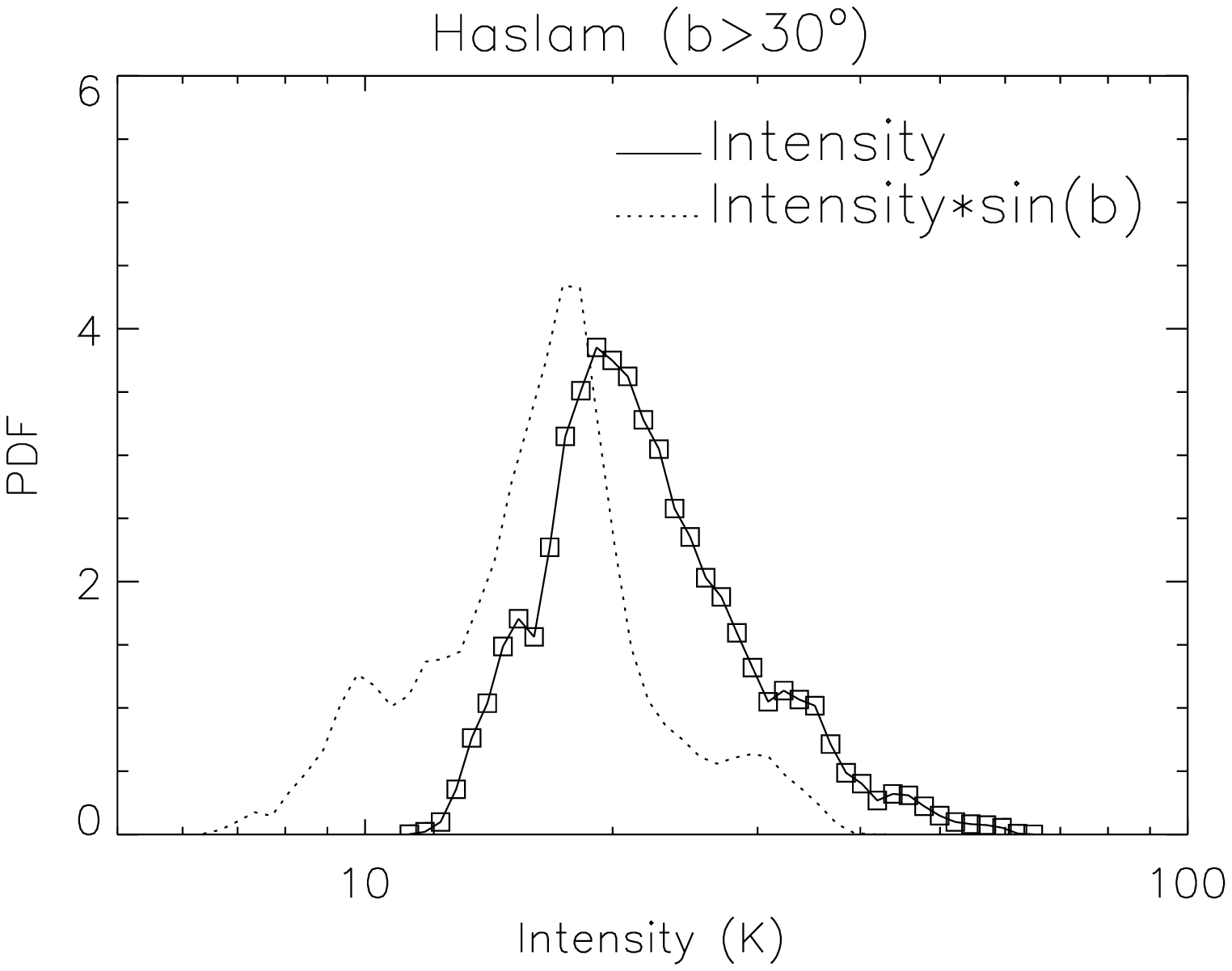}
\caption{ 
   Latitude distribution of average intensity. 
   The solid line shows the usual average intensity. 
   The dashed line and the dotted line depict average taken near
   the Galactic center ($-45^{\circ}\le l_{Gal} \le 45^{\circ}$) 
   and the Galactic anti-center ($135^{\circ}\le l_{Gal} \le 225^{\circ}$),
   respectively. Here $l_{Gal}$ is the galactic longitude.
\label{fig:avg}
}
\caption{ 
   PDF for $|b|>30^\circ$. 
   PDF of synchrotron intensity (solid curve) may be composed of
   two components: One has a peak at $T\sim 20K$ and the other
   at $T\sim 35K$.
   However, 
    that of intensity time $\sin b$ (dotted curve) does not
    seem to be physically meaningful.
\label{fig:pdf}
}
\end{figure*}

Fig.~\ref{fig:avg} shows average intensity as a function of
galactic latitude $b$. We only use data for the Galactic Northern sky.
The synchrotron emission is roughly constant for high galactic latitudes 
($b\gtrsim 30^{\circ}$), as noted by earlier studies 
(e.g.~de Oliveira-Costa et al. 2003).
The usual average intensity (solid line) and the average taken near the 
Galactic center show a sharp rise towards the Galactic plane.
But, the average taken over the Galactic anti-center does not show such a 
sharp increase towards the Galactic plane.
The latitude profile of synchrotron intensity cannot be explained by
a single emission component.
The constancy of synchrotron intensity for $b\gtrsim 30^{\circ}$ implies that
the emission is originated from the Galactic halo.
If it is coming from a thin disk, the intensity should scale as $\sin b$.
If there is only the halo component, we will not have such a sharp increase
of intensity as we approach to $b = 0^{\circ}$.
Therefore, we need to take into account additional thin disk component for
the Galactic plane 
(i.e. $b \sim 0^{\circ}$).

Judging from visual inspection,
the probability density function (PDF) of synchrotron intensity 
for $b  \ge 30^{\circ}$ can be
explained by a combination of two components.
One component has a peak at $T\sim 20K$ and the other has a peak
at $T\sim 35K$. The latter may correspond to the emission from
the North Polar Spur.
The former seems to have
roughly a symmetric shape (Fig.~\ref{fig:pdf}).
However, the PDF of intensity times $\sin b$ shows a more complicated
shape (dotted line).

In this section, we study synchrotron emission from the Galactic halo 
(i.e. $b \gtrsim 30^{\circ}$) and the Galactic disk (i.e. $|b|\le 2^{\circ}$)
separately. 
Our main goal is to see if statistics of synchrotron emission from the halo
is consistent with turbulence models.
When it comes to synchrotron emission from the Galactic disk,
it is not easy to separate diffuse emission and emission from discrete sources.
Therefore, we do not try to study turbulence in the Galactic disk.
Instead, we will try to estimate which kind of emission is dominant
in the Galactic disk.

\subsection{Structure of synchrotron halo}
How are the observed angular spectrum of synchrotron emission and 
the spectrum of 3-dimensional MHD turbulence related?
In order to understand the relation, we need to understand
the structure of the Galactic halo.
There exist several models for the diffuse Galactic radio emission.
Beuermann, Kanbach, \& Berkhuijen (1985) showed that a two-component model,
a thin disk embedded in a thick disk, can explain observed synchrotron 
latitude profile.
They claimed that the equivalent width of the thick disk is 
about several kiloparsecs
and thin disk has approximately the same equivalent width as the gas disk.
They assumed that, in the direction perpendicular to the Galactic plane,
the emissivity $\epsilon$ of each component follows 
\be
   \epsilon(z) = \epsilon(0) sech(z/z_0),
\ee
where $z$ is the distance from the Galactic plane and $z_0$ is a constant.
Recently, several Galactic synchrotron emission models have been proposed
in an effort to separate Galactic components from WMAP polarization data
(see, for example, 
                   Page et al.~2007; 
                   Sun et al.~2008; Miville-Deschenes et al.~2008; 
                   Waelkens et al.~2008).
All the models mentioned above assume the 
existence of a thick disk component with
scale height equal to or greater than $1kpc$.
Sun et al.~(2008) considered an additional local spherical component
motivated by the local excess of the synchrotron emission that might be related to
the ``local bubble''.

The detailed modeling of the Galactic synchrotron emission is beyond the 
scope of our paper. 
We will simply assume that there is a thick component with a scale height of
$\sim 1kpc$. We will also assume that 
there could be an additional local spherical component.
Then, what will be the relation between spectrum of 3-dimensional turbulence 
and the observed angular spectrum of synchrotron emission?



\begin{figure*}[h!t]
\includegraphics[width=0.32\textwidth]{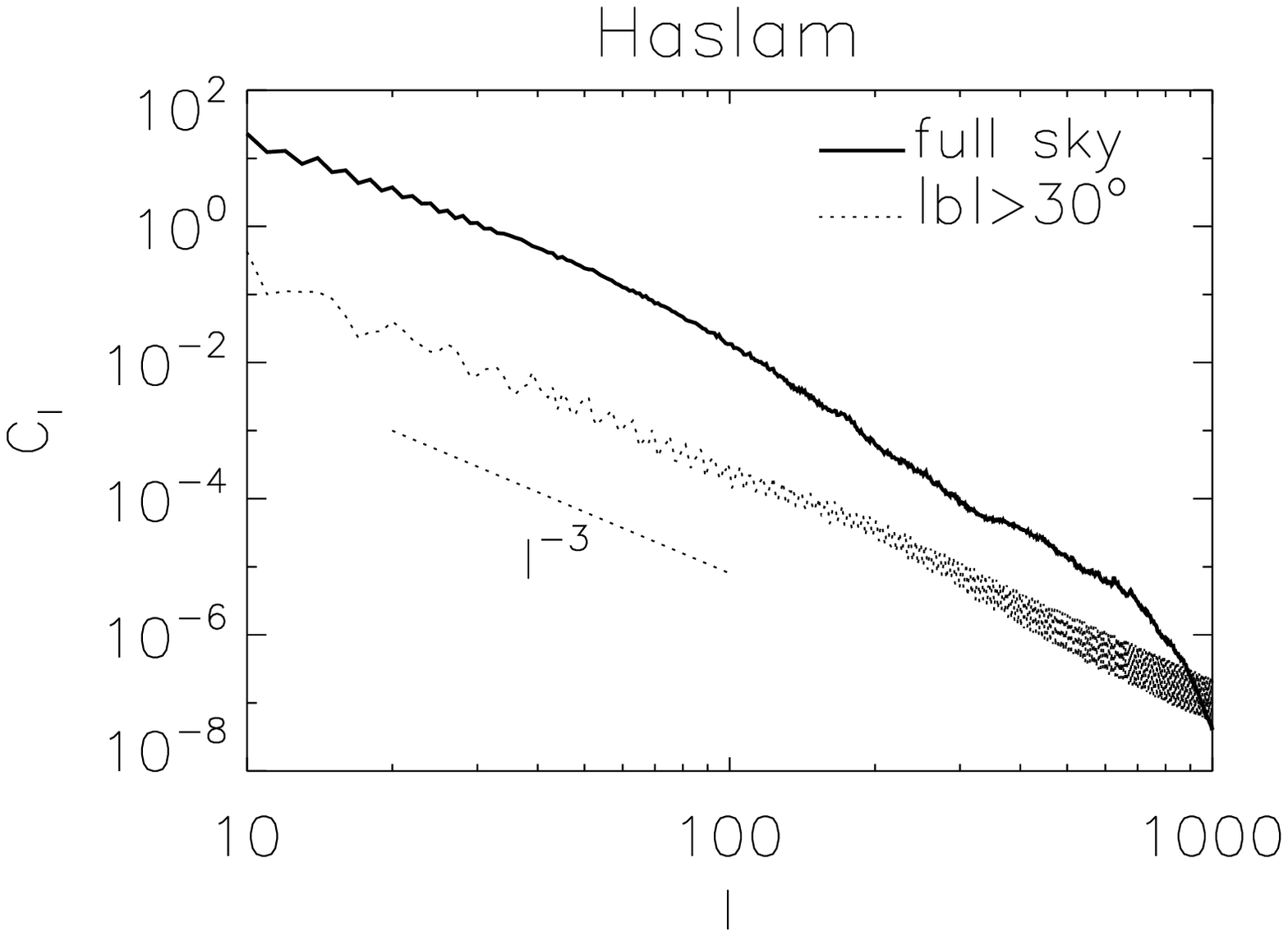}  
\includegraphics[width=0.32\textwidth]{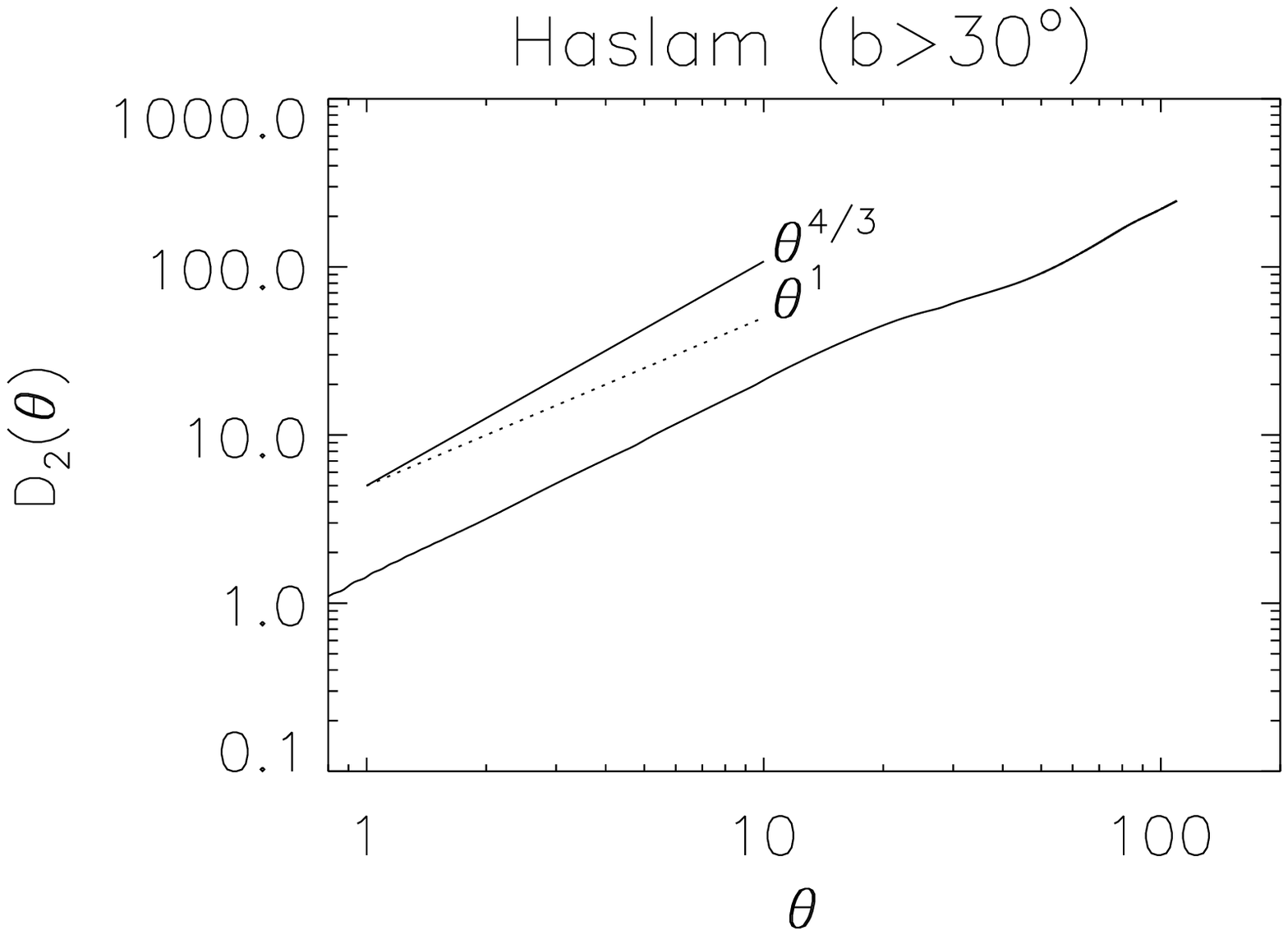}   
\includegraphics[width=0.32\textwidth]{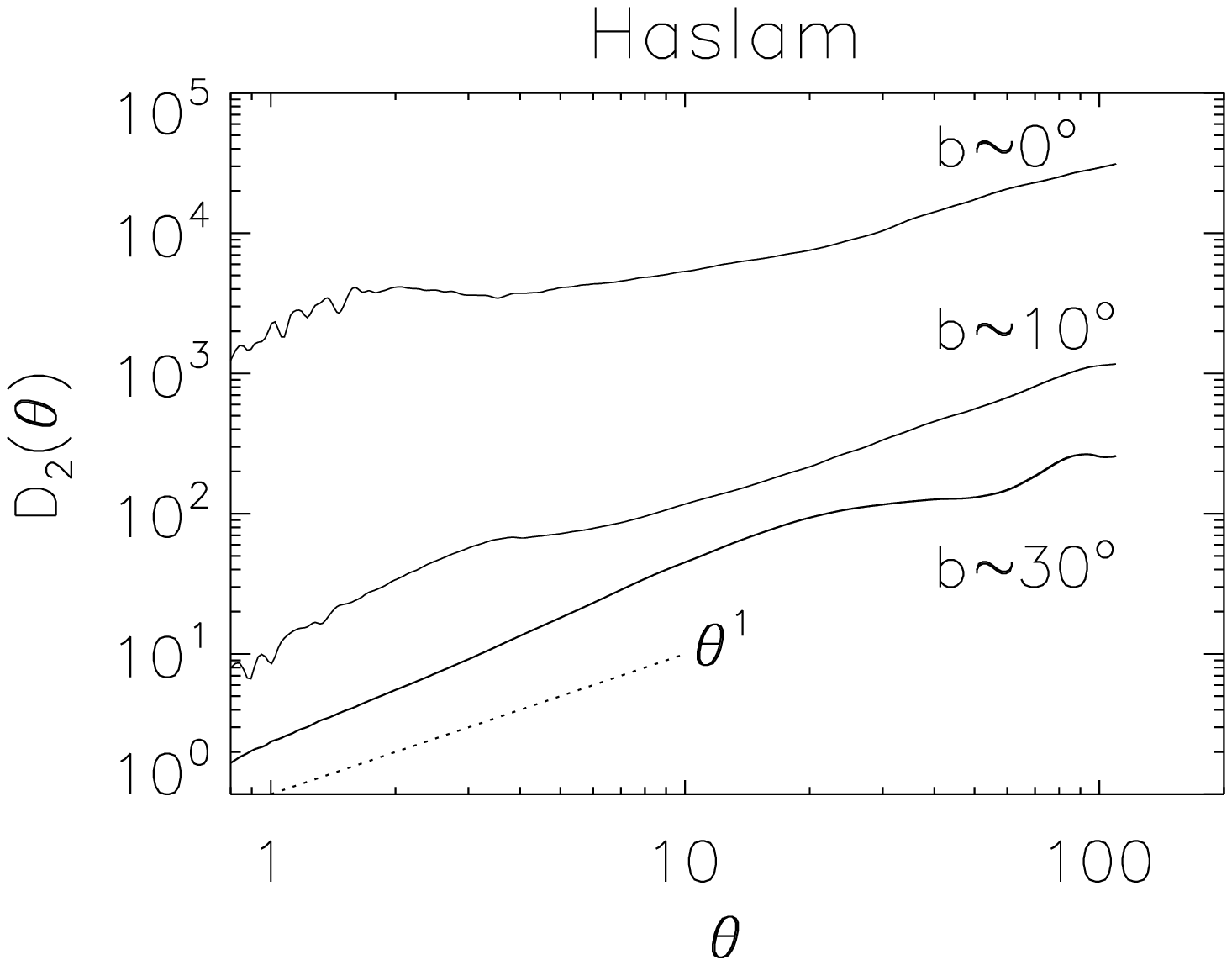}   
\caption{ 
   Haslam 408MHz map.
  {\it Left}: Angular spectra of full-sky and partial-sky ($|b|>30^{\circ}$) maps.
   Spectrum of partial-sky map (thin curve) suffers from edge effect because 
   pixels with $|b|\le 30^{\circ}$ are set to zero.
 {\it Middle}: Second-order angular 
   structure function shows a slope of $\sim 1.2$, which
   is shallower than that of Kolmogorov turbulence (slope $= 5/3$).
  {\it Right}: Structure function as a function of galactic latitude.
    {}From bottom to top, the second-order angular 
     structure functions are obtained
     for thin stripes ($|\Delta b| \le 2^\circ$) along 
     galactic latitudes of $30^{\circ}$,
      $10^{\circ}$, and $0^{\circ}$.
}
\label{fig:sfetc-has}
\end{figure*}

\subsection{Spectrum and Structure function of the 408MHz Haslam map}
In left panel of Fig.~\ref{fig:sfetc-has} we plot angular power spectrum
of synchrotron emission at 408Mhz.
The upper curve (solid curve) is the angular spectrum of the all-sky data.
The lower curve (dotted curve) is that of high galactic latitude.
We obtain the latter as follows. We 
first set the synchrotron intensity to zero
for pixels with $|b| \le 30^{\circ}$ and then calculate 
the angular spectrum
using {\bf anafast} package.\footnote{
  HEALPix Homepage: http://healpix.jpl.nasa.gov/ }
Due to the partial sky coverage, the spectrum (lower curve)
show an oscillatory behavior.
The slope of the lower curve is very close to that of the 
straight line, which has a slope of -3.
This result is consistent with earlier results for the 408-MHz Haslam map
(Tegmark \& Efstathiou 1996; Bouchet, Gispert, \& Puget 1996).
Recently La Porta et al. (2008) performed a 
comprehensive angular power spectrum
analysis of all-sky total intensity maps at 408MHz and 1420MHz. 
They found that the slope is close to -3 for high galactic latitude regions.
Other results also show slopes close to -3. For example, 
using Rhodes/HartRAO data at 2326 MHz (Jonas, Baart, \&
Nicolson 1998), Giardino et al. (2001b) obtained a $\sim 2.92$ 
for high galactic latitude regions with
$|b| > 20^\circ$. Giardino et al. (2001a) obtained a $\sim 3.15$ for high
galactic latitude regions with $|b| > 20^\circ$ from the Reich \& Reich
(1986) survey at 1420 MHz. Bouchet \& Gispert (1999) 
also obtained a $\sim l^{-3}$
spectrum from the 1420 MHz map.
%
%

Since we obtained angular spectrum from data with incomplete sky coverage,
the slope of the spectrum may be contaminated by the edge effect.
To avoid the effect, we may calculate angular correlation function $K(\theta)$
 first and 
then obtain $C_l$ using Eq.~(\ref{eq:k2cl}).
But the angular spectrum $C_l$ obtained in this way is, in general, noisy.
Therefore it is not easy to accurately measure the slope of the spectrum.
When we want to obtain only the slope of the angular spectrum
on small angular scales, hence the
slope of the 3D spatial turbulence spectrum,
we can use the second-order angular structure function:
\be
   D_2 (\theta) 
   = < | T({\bf e}_1) -T({\bf e}_2) |^2 >. 
\ee
As we discussed in \S\ref{sect:smangle}, 
the slope of the second-order structure function will tell us 
about the slope of the angular spectrum \footnote{
 One should be careful if there is a white noise.
 In the presence of white noise, second-order structure function will be
 $D_2 (\theta) 
   = < | T({\bf e}_1) +\delta_1 -T({\bf e}_2)-\delta_2 |^2 >
  = < | T({\bf e}_1) -T({\bf e}_2) |^2 > + < |\delta_1-\delta_2|^2 >,
  $
where $\delta_1$ and $\delta_2$ represent noise (Chepurnov, private communication).
Therefore, the noise can interfere accurate measurement of the slope.
However, if $\lim_{\theta \rightarrow 0}D_2$ is small enough, we can ignore the noise.
This is true in our case.
The Haslam 408MHz map shows $D_2(0.015^\circ) \sim 0.05$, which is sufficiently smaller
than values $D_2$ shown in Fig.~\ref{fig:sfetc-has}.
}.

In the middle panel of Fig.~\ref{fig:sfetc-has} we show the second-order
structure function for the Galactic halo (i.e. $|b| > 30^{\circ}$).
The slope of the second-order structure function lies between
those of two straight lines.
The steeper line has a slope of 4/3 and the other one has a slope of 1.
The actual measured slope is $\sim 1.2$.
This result implies that the 3D turbulence spectrum is $E_{3D}(k)\propto k^{-3.2}$,
which is shallower than the Kolmogorov spectrum $E_{3D}(k)\propto k^{-11/3}$.

\begin{figure*}[h!t]
\includegraphics[width=0.40\textwidth]{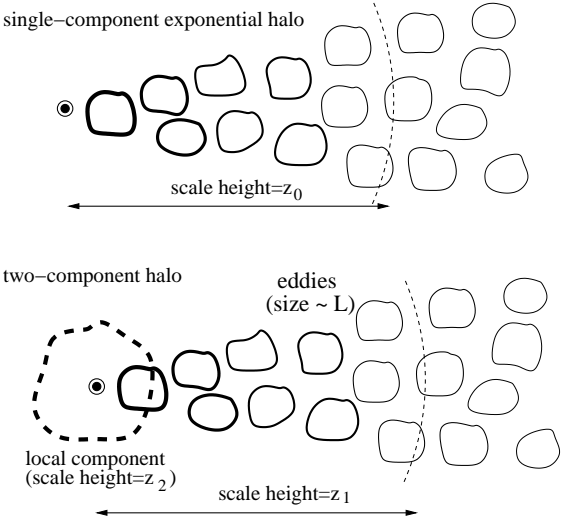}  
\caption{Halo models with stratification.
  {\it Upper plot}:
  Exponentially stratified halo.
  We take $z_0 (=r_0) =1kpc$ and $L$ (=eddy size) $=100pc$.
{\it Lower plot}: Two-component halo.
  We take  $z_1 (=r_1) =1kpc$, $z_2 (=r_2) =100pc$,  
  and $L$ (=eddy size) $=100pc$.
}
\label{fig:2models}
\end{figure*}

\subsection{Effects of inhomogeneity and model calculations} 
\label{sect:3models}
Then why is the slope shallower than that of Kolmogorov?
Chepurnov (1999) gave a discussion 
about the effects of density stratification on the slope.
He used a Gaussian disk model and semi-analytically showed that
the slope of the angular spectrum can be shallower than
that of Kolmogorov.
In this subsection, we present further discussions.

In the model discussed in \S\ref{sect:spsfco}, 
we assumed turbulence is homogeneous.
However, this is certainly an unrealistic assumption for the Galactic halo.
Synchrotron emission models (see previous subsection) assume either exponential 
($e^{-z/z_0}$) or square of hyperbolic secant ($sech^2[z/z_0]$) law
for synchrotron emissivity, where $z$ is the distance from the 
Galactic plane.
For simplicity, we assume the observer is at the center of
a spherical halo. That is, the geometry is not plane-parallel, but spherical.
In what follows, we use $r$, instead of $z$, to denote the 
distance to a point.

To illustrate the effects of this inhomogeneity, we test 3 models:
\begin{enumerate}
\item {\bf Homogeneous halo}: Turbulence in halo, thus emissivity, is homogeneous. 
      Turbulence has a sharp
      boundary at $d_{max}=1kpc$. The outer scale of turbulence is 100pc.
      Basically, this model is the same as the one we considered
      in \S\ref{sect:spsfco}.   
\item {\bf Exponentially stratified halo}: Emissivity shows an exponential decrease, 
      $\epsilon(r) \propto e^{-r/r_0}$.
      We assume $r_0 = 1kpc$ and the halo truncates at $r=8kpc$.
      The outer scale of turbulence is 100pc. See Fig.~\ref{fig:2models}.
\item {\bf Two-component halo}: Emissivity decrease as 
      $\epsilon(r) \propto \epsilon_1 e^{-r/r_1}+\epsilon_2 e^{-r/r_2}$, where
      $\epsilon_2=10\epsilon_1$, $r_1 = 1kpc$, and $r_2=100pc$.
      The halo truncates at $r=8kpc$.
      The outer scale of turbulence is 100pc.
      The second component mimics local enhancement of synchrotron emissivity. See Fig.~\ref{fig:2models}.
\end{enumerate}

\begin{figure*}[h!t]
\includegraphics[width=0.32\textwidth]{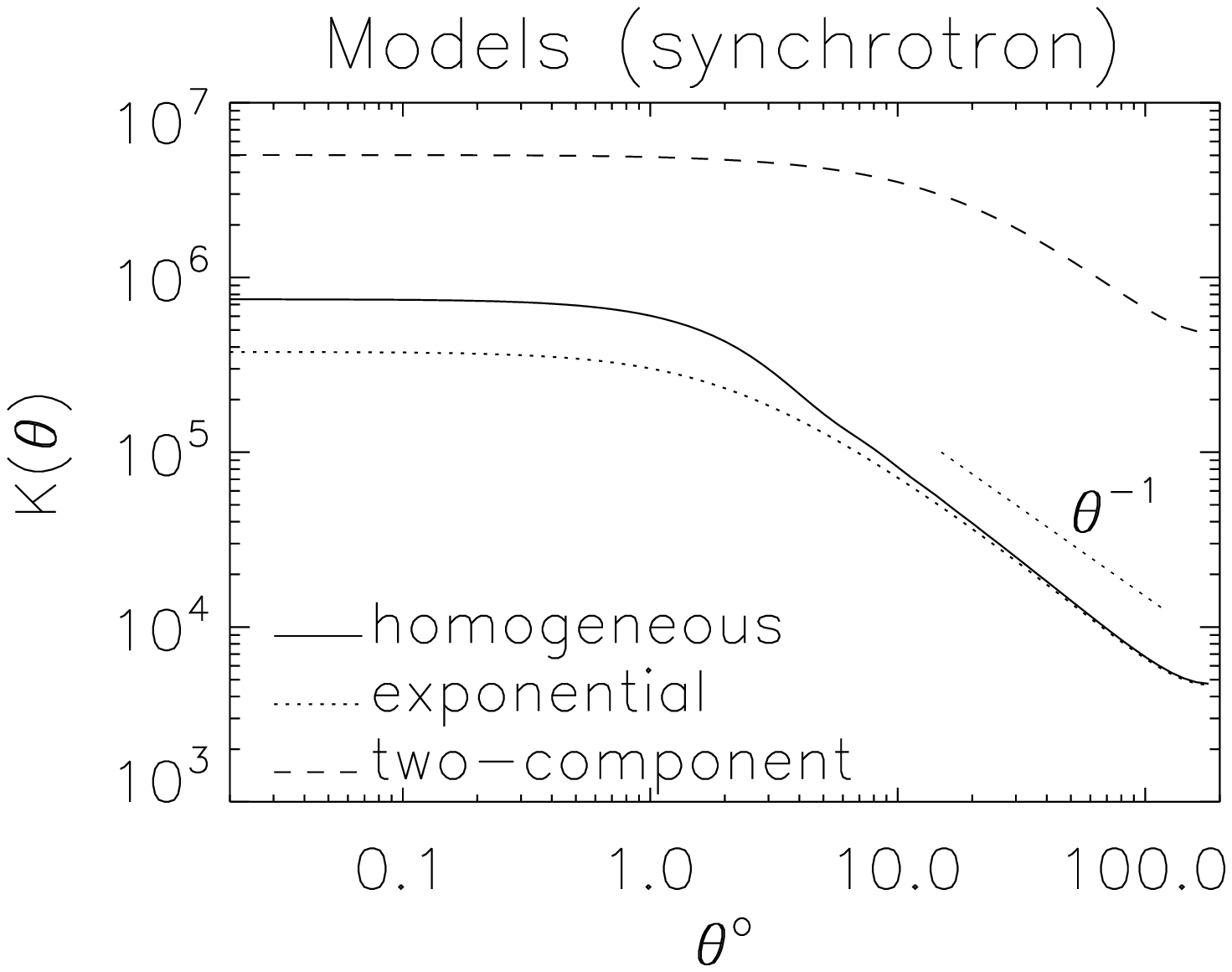}  
\includegraphics[width=0.32\textwidth]{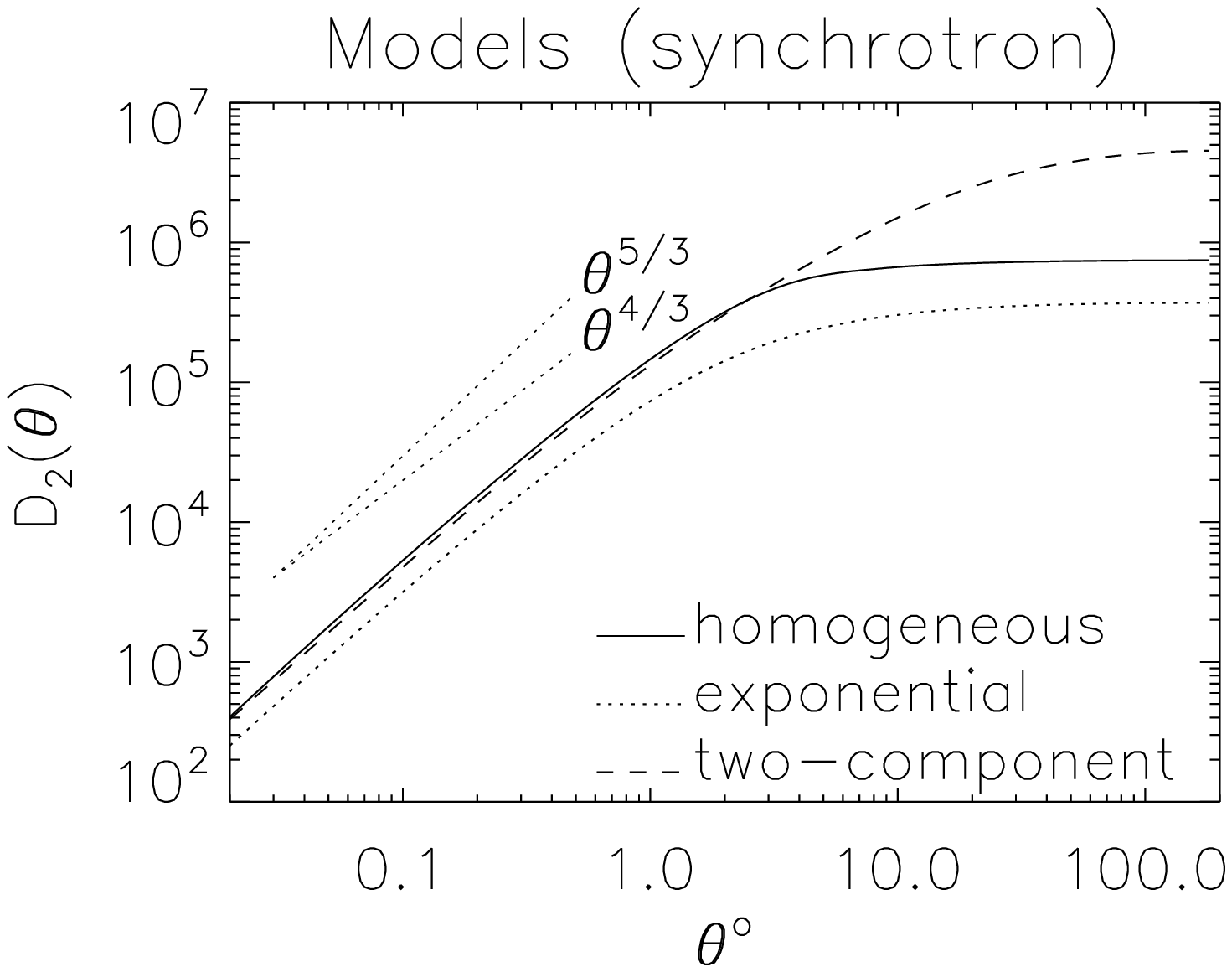}   
\includegraphics[width=0.32\textwidth]{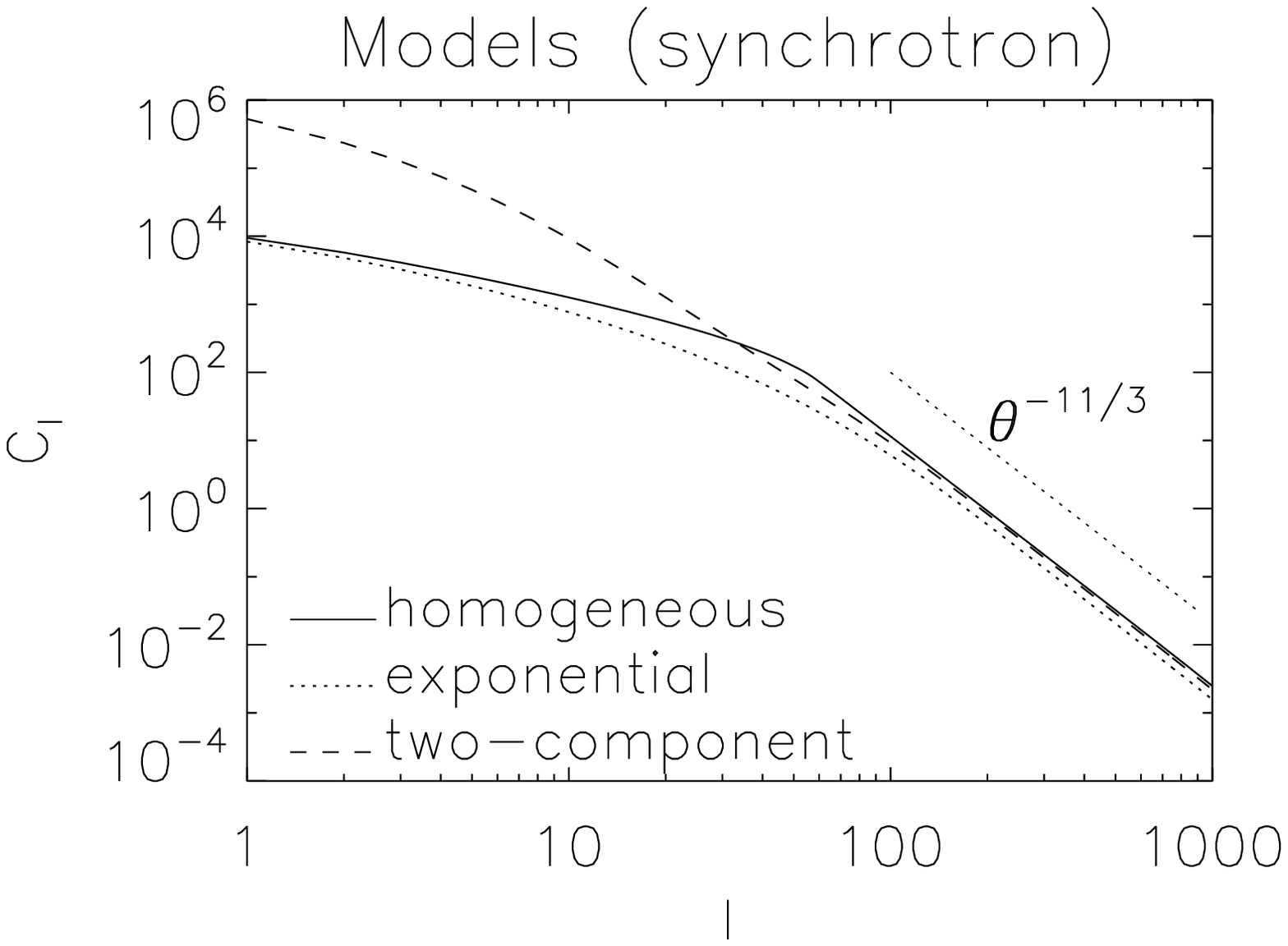}  
\caption{ 
   Model calculations. 
   Three toy models for emissivity profiles in the Galactic halo are considered:
   homogeneous (solid line) halo, exponentially stratified halo (dotted line),
   and two-component exponential halo (dashed line).
      {\it Left}: Angular correlation functions.
                     When angular separation is large, angular correlation
                     functions follow the universal relation:
                         $(\pi -\theta)/\sin\theta \sim 1/\theta$.
      {\it Middle}: Second-order structure functions.
                    When angular separation is small,
                    the slope of the homogeneous turbulence (solid line) 
                    is compatible with $5/3$.
                    But those of 
                    stratified halo models are shallower.
                    The two-component model (dashed line) shows a slope
                    compatible with $-3$, if we measure average slope
                    between $\theta \sim 0.5^\circ$ and $\sim 10^\circ$.
      {\it Right}: Angular spectra.
                   In general, the spectra are compatible with the 3D spatial
                   turbulence spectrum of $l^{-11/3}$ for large $l$.
                   The stratified halo models show 
                   shallower slopes, if we measure average slope
                   between $l\sim 10$ and $\sim 200$ for instance.
                   The homogeneous turbulence model also gives a
                   shallower slope for these values of $l$.
                   But, its spectrum shows a easily noticeable break
                   near $l\sim 50$.
}
\label{fig:model}
\end{figure*}

We numerically calculate the angular correlation function $K(\theta)$ and
the second-order structure $D_2(\theta)$ from
\bea
    K(\theta)=\int dr_1 \int dr_2 ~{\cal K}(|{\bf r}_1-{\bf r}_2|)
    \epsilon(r_1) \epsilon(r_2),   \label{eq_ctheta}  \\
    D_2(\theta) \propto \overline{T} - K(\theta),
\eea
where $|{\bf r}_1-{\bf r}_2|=r_1^2+r_2^2-2r_1r_2 \cos{\theta}$,
$\epsilon(r)$ is the synchrotron emissivity, 
$\overline{T}=\lim_{\theta \rightarrow 0} K(\theta)$,
and 
we use the spatial correlation function ${\cal K}(r)$ obtained from the relation:
\be
 {\cal K}(r) \propto \int_{0}^{\infty} 4\pi k^2 E_{3D}(k) \frac{ \sin kr }{ kr } dk
               \label{c_r_3D}
\ee
where the spatial spectrum of emissivity $E_{1D}$ has the form:
\be
 E_{3D}(k) \propto \left\{ \begin{array}{ll} 
                              \mbox{constant}   & \mbox{if $k\le k_0$} \\
                              (k/k_0)^{-11/3}   & \mbox{if $k\ge k_0$,}
                      \end{array}
              \right. \label{E_3D}
\ee
which is the same as Kolmogorov spectrum for $k\ge k_0$ ($ \sim 1/L$).
The reason we use  a constant spectrum for $k\le k_0$ is explained in Appendix B
(see also Chepurnov 1999).
We obtain the angular spectrum from the relation:
\be
   C_l \propto \int P_l(\cos{\theta}) K(\cos \theta)\ d(\cos\theta), \label{eq:k2cl}
\ee
where $P_l$ is the Legendre polynomial.

In Fig.~\ref{fig:model}, we plot the calculation results.
The angular correlation function $K(\theta)$ does not change much 
when $\theta$ is small, and
follows $\sim (\pi-\theta)/\sin\theta \sim 1/\theta$ when $\theta$ is large.
The critical angle is a few degrees for homogeneous model (thick solid curve)
and single-component exponential model (dotted curve).
As we discussed earlier, the critical angle for homogeneous turbulence is
$\sim (L/d_{max})^{rad}\sim 6^\circ$, where $d_{max}$ ($=1kpc$ in our model) 
is the distance to the farthest eddy.
In Fig.~\ref{fig:model} (left panel) we clearly see that
the slope of $K(\theta)$ changes near $\theta \sim 6^\circ$.
The second-order structure function $D_2(\theta)$ 
also shows a change of slope near the
same critical angle ($\theta\sim 6^\circ$).
In single-component exponential model (dotted curve), 
the value of $d_{max}$ is not important.
Instead, the scale height $z_0$ is a more important quantity, 
which is $1kpc$ in our model.
In left and middle panels of Fig.~\ref{fig:model}, we observe
 that the 
single-component exponential model also show a change of slope 
near $\theta \sim$ a few degrees. 
Therefore, we can interpret that the critical angle
for stratified turbulence is $\sim L/z_0$, instead of $\sim L/d_{max}$

Then, can we answer our earlier question of why
the observed slope is shallower than that of Kolmogorov?
Let us take a look at the right panel of Fig.~\ref{fig:model}.
All 3 models show that the slope of $c_l$ is almost Kolmogorov one for
$l\gtrsim l_{cr}\sim \pi d_{max}/L \sim 30$.
However, if we measure average slope of $c_l$ between $l=10$ and $200$,
we obtain slopes shallower than Kolmogorov.
The single-component model and the homogeneous model
give similar average slopes, which is a bit shallower than $-3$.
However, the homogeneous model
gives a more abrupt change of slope near $l\sim 30$.
(In fact, right panel of Fig.~\ref{fig:model} shows the break happens 
 near $l\sim 50$.)
The average slope of 
the two-component model gives more or less gradual change of the slope and
the slope is very close to $-3$ for a broad range of multipoles $l$.
It is difficult to tell which model is better because the models are
highly simplified.
But, the two-component model looks the most promising. 

Note that, compared with the two-component model, 
the single-component model shows a more or less sudden change of slope
near $l\sim l_{cr} \sim \pi z_0/L \sim 30$.
Therefore, if the single-component model is correct, the scale height $z_0$ cannot be
much larger than $\sim 10$ times the outer scale of turbulence $L$.
If $z_0$ is much larger than $\sim 10L$, $l_{cr}$ becomes smaller and
we will have almost Kolmogorov slope for $l\gtrsim 10$.
We also note that it is possible that
 3D spatial turbulence spectrum itself can be shallower than
the Kolmogorov one. 
That is, it is possible that 
spectrum of ${\bf B}({\bf r})$, hence that of $B^2({\bf r})$, can be
shallower than the Kolmogorov one.
For example, recent studies show that strong MHD turbulence can have a $k^{-3.5}$
spectrum, rather than $k^{-11/3}$ (Boldyrev 2006; Beresnyak \& Lazarian 2006).
If this is the case, the observed angular spectrum can be shallower
than Kolmogorov for $l>l_{cr}$.

\subsection{Synchrotron emission from Galactic disk}

In right panel of Fig.~\ref{fig:sfetc-has} we show how the second-order structure
function changes with galactic latitude.
The lower curve is the second-order angular structure function 
obtained from pixels in the range of $28^{\circ}\le b \le 32^{\circ}$.
The middle and upper curves are the second-order 
angular structure functions
obtained from pixels in the range of $8^{\circ}\le b \le 12^{\circ}$ and
$-2^{\circ}\le b \le 2^{\circ}$, respectively.
The middle and upper curves clearly show break of slopes near
$\theta \sim 3^\circ$ and $\sim 1.5^\circ$, respectively.
When the angular separation is larger than the angle of the break,
the structure function becomes almost flat.
As we move towards the Galactic plane, the sudden changes of slopes happen
at smaller angles.

What causes this break of slope?
There are at least two possibilities.
First, a geometric effect can cause it. 
As we discussed in \S\ref{sect:spsfco}, change of slope occurs near
$\theta_c \sim L/d_{max}$.
As we move towards the Galactic plane, the distance to the farthest eddy, $d_{max}$,
will increase. As a result, the critical angle $\theta_c \sim L/d_{max}$ will
decrease.
Therefore, we will have smaller $\theta_c$ towards the Galactic plane.
 This may be what we observe in the right panel of Fig.~\ref{fig:sfetc-has}.
Second, discrete synchrotron sources can cause flattening of the structure
function on angular scales larger than their sizes. 
Although the map we use was reprocessed to remove
strong point sources, there might be unremoved discrete sources.
When filamentary discrete sources dominate synchrotron emission, the 
second-order structure function
will be flat on scales larger than the typical width of the sources.
In reality, both effects may work together.
At this moment, it is not easy to determine which effect is more important.

\subsection{On the polarized synchrotron emission}
Roughly speaking, the shape of
the angular spectrum of polarized synchrotron emission will 
be similar to that of the total intensity at {\it mm} wavelengths.
However, at longer wavelengths, Faraday rotation and depolarization effects
 may cause flattening of the angular spectrum, which has been
actually reported
(see de Oliveira-Costa et al. 2003 and references therein).
On the other hand, La Porta et al. (2006) analyzed the new DRAO 1.4GHz 
polarization survey and obtained angular power spectra with
power-law slopes in the range $[-3.0, -2.5]$.
More observations on the polarized synchrotron foreground emission
can be found in Ponthieu et al. (2005),
Giardino et al. (2002), Tucci et al. (2002), Baccigalupi et al. (2001).
In this paper, we do not discuss further about
polarized synchrotron emission.
Readers may refer to recent models about the polarized synchrotron emission
(Page et al. 2007; Sun et al. 2008; Miville-Deschenes et al. 2008;
 Waelkens et al. 2008).

\section{Properties of the Dust Emission Map} \label{sect:dust_em}

Thermal dust emission is also an important ingredient of CMB foregrounds.
In this section, we analyze a model dust emission map created by
Finkbeiner et al. (1999),
which is 
available at the NASA LAMBDA website (http://lambda.gsfc.nasa.gov/).

\begin{figure*}[h!t]
\includegraphics[width=0.32\textwidth]{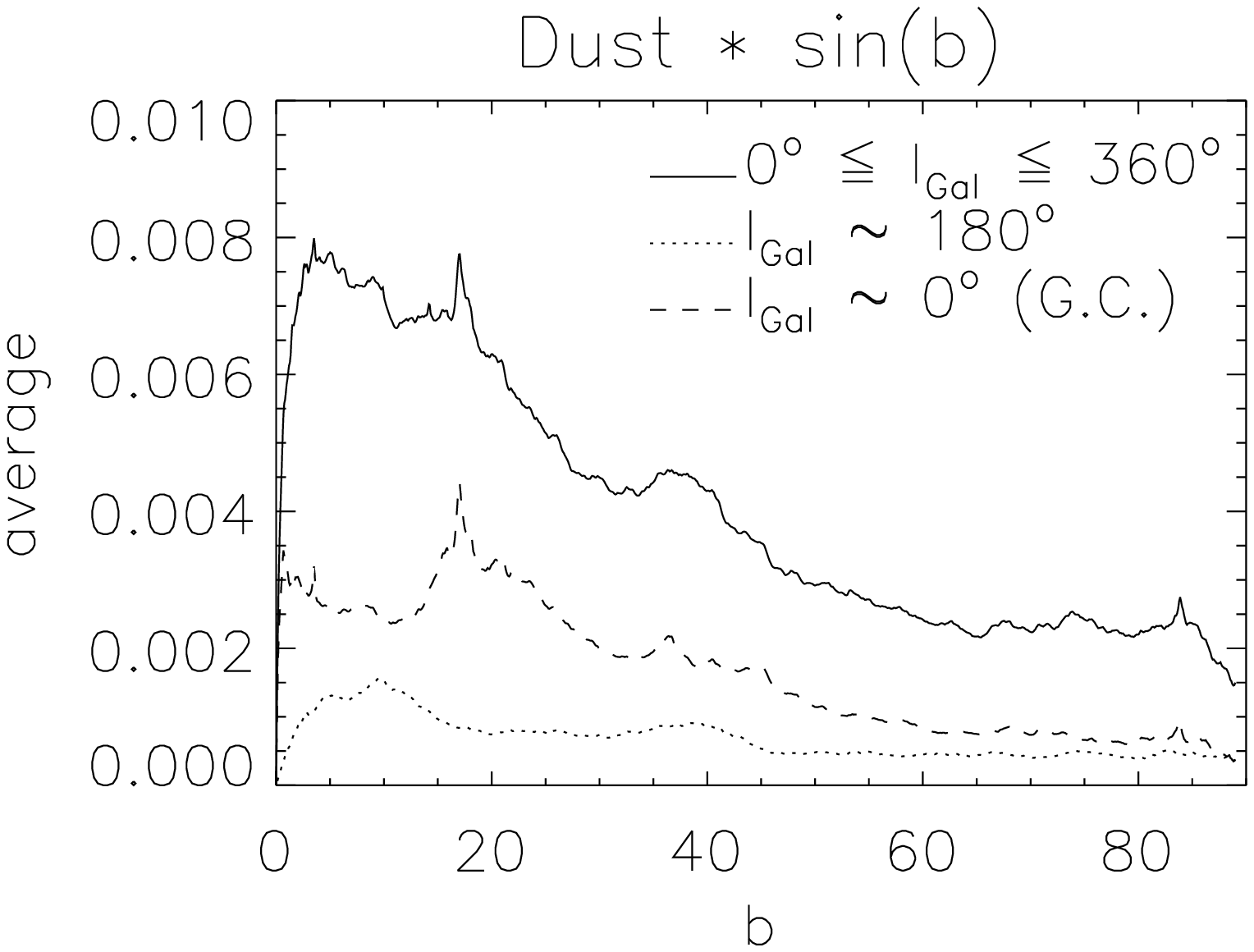}    
\includegraphics[width=0.32\textwidth]{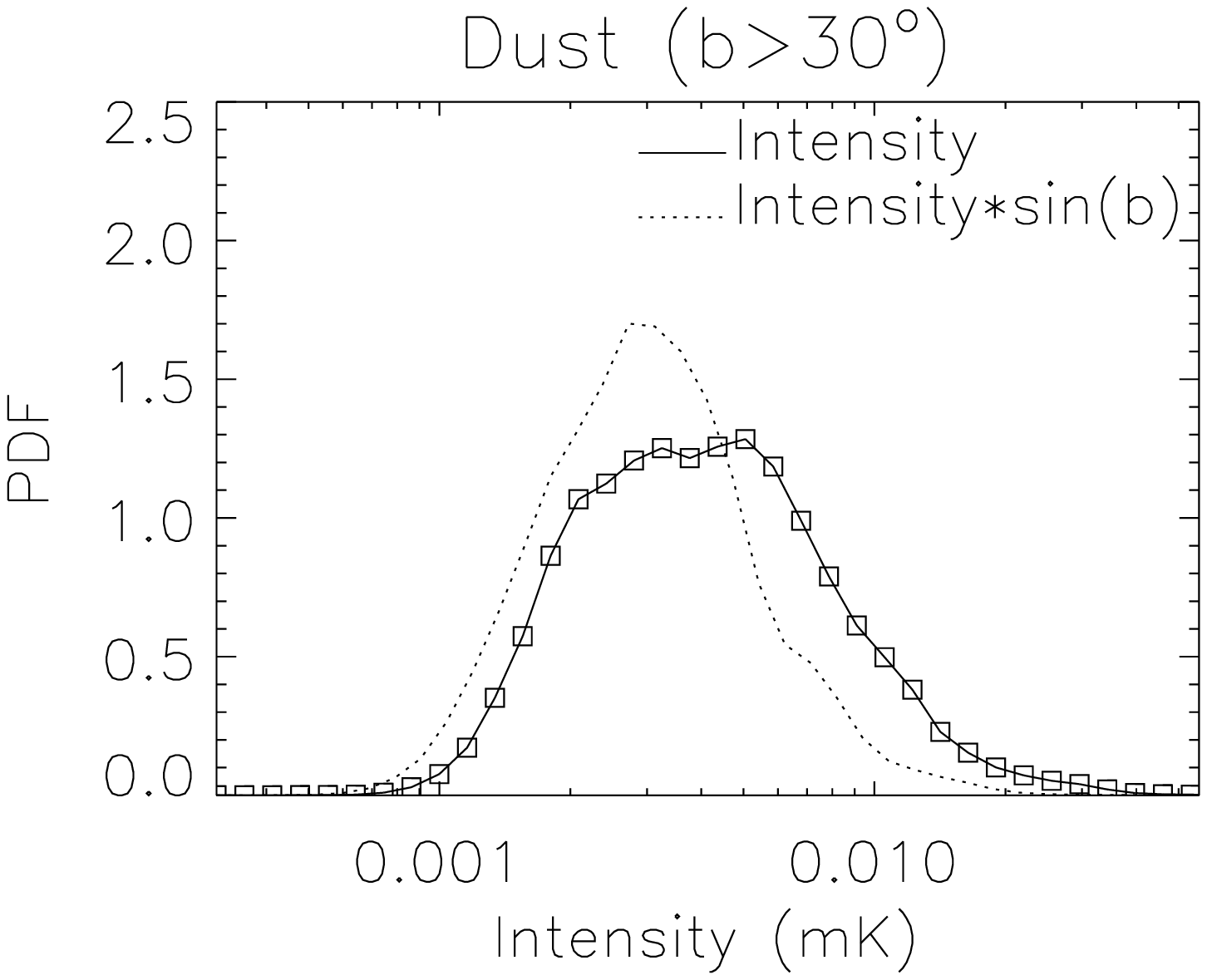}  
\includegraphics[width=0.32\textwidth]{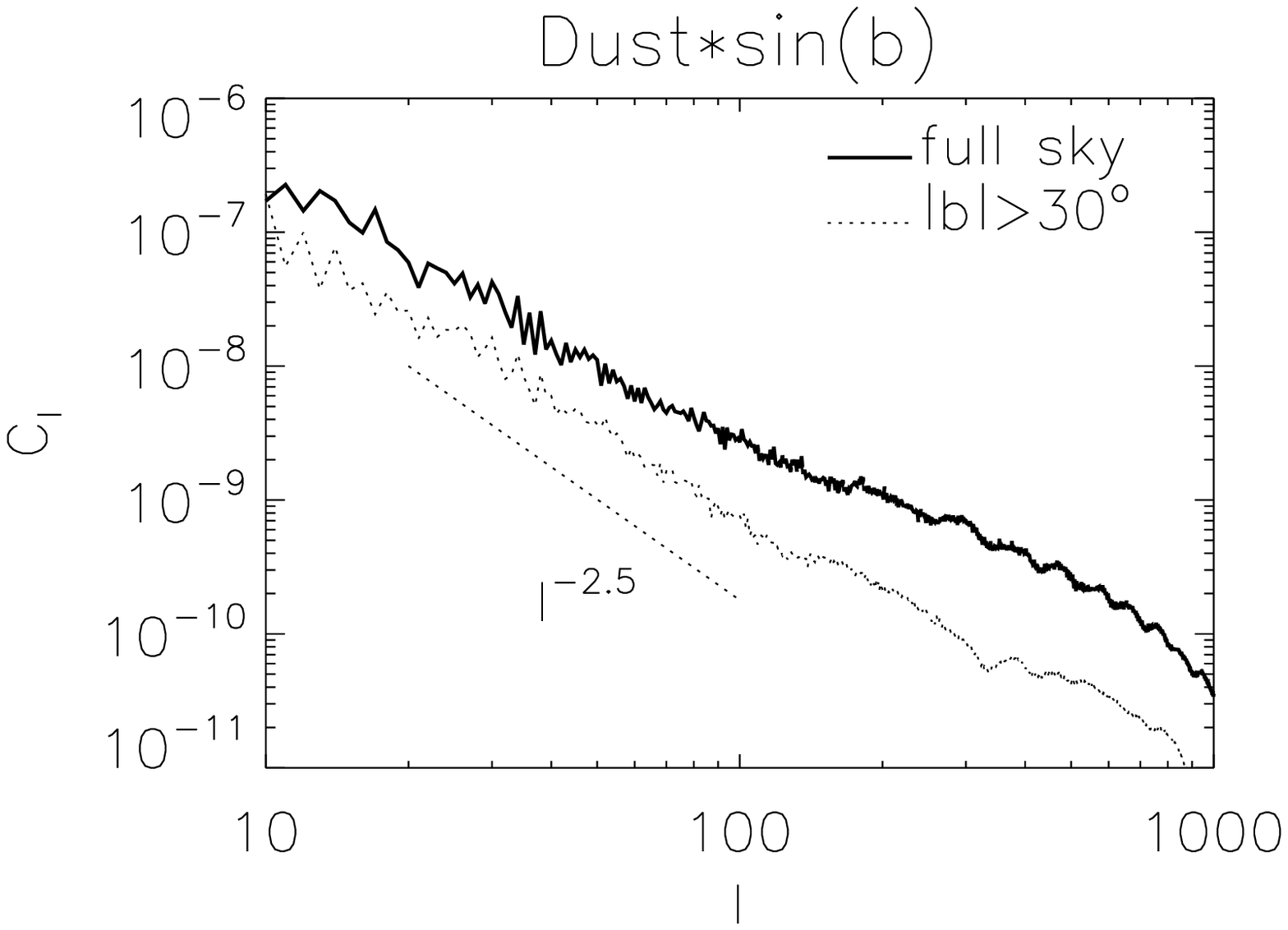}  
\caption{ 
   Average dust emission intensity times $\sin b$. 
  {\it Left}: Latitude profile of averaged dust emission 
     intensity time $\sin b$. 
     Average value is roughly constant in high latitude region.
     The solid line represents the average taken over $360^{\circ}$. 
   The dashed line and the dotted line depict average taken near
   the Galactic center ($-45^{\circ}\le l_{Gal} \le 45^{\circ}$) 
   and the Galactic anti-center ($135^{\circ}\le l_{Gal} \le 225^{\circ}$),
   respectively.
 {\it Middle}: 
    PDF for $|b|>30^\circ$.  PDF of dust intensity times $\sin b$ 
   (dotted curve) shows a rough symmetry, but
    that of dust intensity (solid curve) does not.
  {\it Right}: Angular spectra of full-sky and partial-sky ($|b|>30^{\circ}$) maps.
    Spectra are for dust map times $\sin b$.
}
\label{fig:dust}
\end{figure*}

In Fig.~\ref{fig:dust}, we present statistical properties of the map.
The map shows rough constancy of emission
for high galactic latitude region when
multiplied by $\sin b$ (left panel of Fig.~\ref{fig:dust}).
The $\sin b$ factor also appears in the PDF: the $\sin b$ factor makes the PDF more
symmetric (middle panel of Fig.~\ref{fig:dust}).
Therefore it is natural to conclude that a thin disk component
dominates the dust map.

Angular spectrum of the dust map shows a slope much flatter than
that of the Haslam map (right panel of Fig.~\ref{fig:dust}).
As in the Haslam map, the upper curve (solid curve) is the angular spectrum of the all-sky data.
The lower curve (dotted curve) is that of high galactic latitude.
The slope of the dotted curve is very close to $-2.5$.
However, due to the edge effect, it is not clear whether or not
the slope is the true one.
Since it is not affected by the 
edge effect, the second-order structure function can reveal the
true scaling relation on small angular scales.
Note that the $l^{-2.5}$ spectrum implies that
the second-order structure function has a slope very close to 
$0.5$.
Indeed, the actual second-order structure function 
of the dust map (see \S\ref{sect:h_o}) shows
a slope of $\sim 0.6$, which corresponds to angular spectrum of $\sim l^{-2.6}$.

The slope of the angular power spectrum of the model dust emission map
is very similar to that of the original FIR data.
Schlegel et al.~(1998) found a slope of $-2.5$ for the original FIR
data. On the other hand, other researchers found
slopes close to $-3$ from other observations 
(see Tegmark et al.~2000 and references therein;
see also Masi et al.~2001).

\begin{figure*}[h!t]
\includegraphics[width=0.45\textwidth]{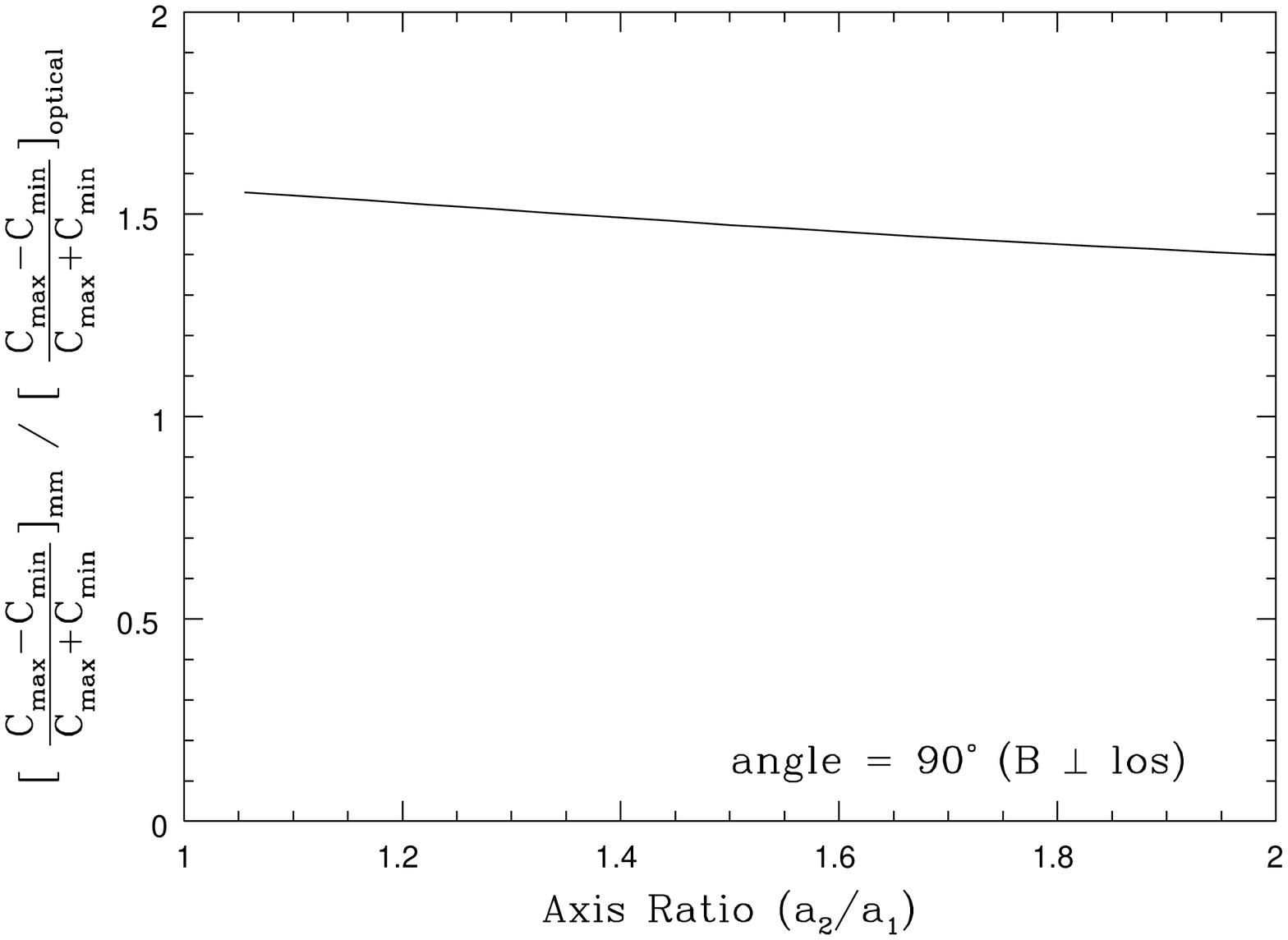}  
\includegraphics[width=0.45\textwidth]{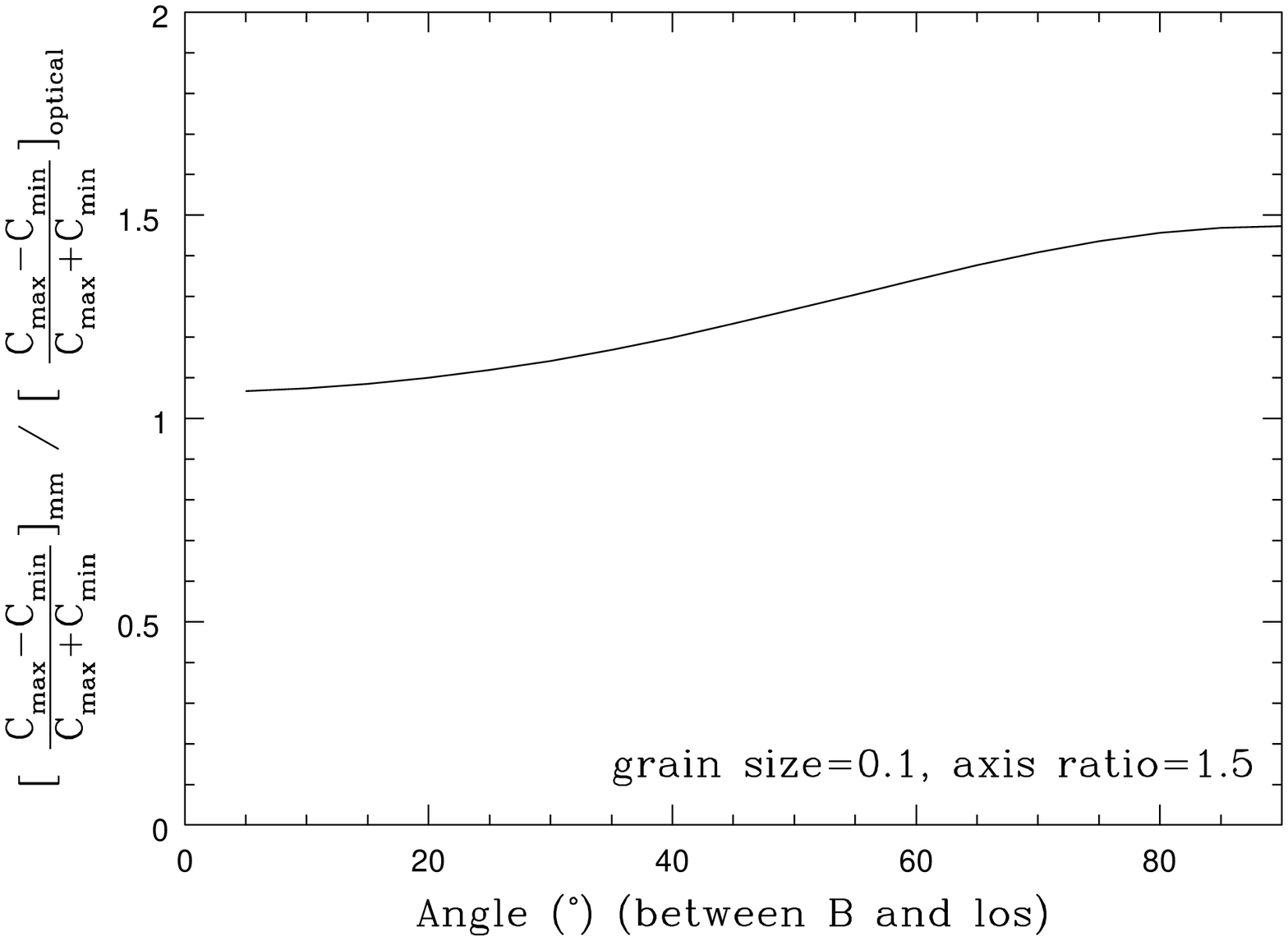}  
\caption{ 
   The ratio of $P_{em,mm}/P_{em,optical}$.
  {\it Left}: The polarization ratio vs.~the axis ratio ($a_2/a_1$) 
   of oblate spheroidal grains.
   The polarization ratio shows only a weak dependence on the axis ratio.
   We assume that the grain size is 0.1$\mu$m, magnetic field is perpendicular
   to the line-of-sight, $\lambda_{optical}=0.5\mu$m,
   and $\lambda_{mm}=1000\mu$m.
  {\it Right}: The polarization ratio vs.~the angle between magnetic field and 
   the plane of the sky.
   We assume that the grain size is 0.1$\mu$m, the grain axis ratio is 1.5,
   $\lambda_{optical}=0.5\mu$m,
   and $\lambda_{mm}=1000\mu$m.
}
\label{fig:ratios}
\end{figure*}

\section{Polarized Emission from Dust}
Polarized radiation from dust is an important
component of Galactic foreground that strongly interferes with
intended CMB polarization measurements (see Lazarian \& Prunet 2001).
Then, what will be the angular spectrum of the polarized radiation from the
foreground dust?
One of the possible ways to estimate the polarized radiation 
{}from dust at the microwave
range is to measure star-light polarization and use
the standard formulae 
(see, for example, Hildebrand et al. 1999) 
relating polarization at different wavelengths.
In this section, we describe how we can obtain a map of 
polarized emission from thermal dust and we discuss
the angular spectrum of the polarized emission from
thermal dust in the high galactic latitude (say, $|b| \gtrsim 20^\circ$).

\subsection{Getting a map of polarized emission from thermal dust}
In principle we can construct 
a polarized dust emission map at {\it mm} wavelengths ($I_{pol,mm}(l,b)$) from 
a dust total emission map ($I_{mm}(l,b)$) 
and a degree-of-polarization map ($P_{em,mm}(l,b)$) at {\it mm} wavelengths:
\begin{equation}
  I_{pol,mm}(l,b)= P_{em,mm}(l,b)~ I_{mm}(l,b),  \label{eq:IIP}
\label{eq_Ipol_mm}
\end{equation}
where $(l,b)$ denotes the galactic coordinate.
However, neither $I_{pol,mm}(l,b)$ nor $I_{mm}(l,b)$ is directly
available.
Therefore, we need indirect methods to get 
$I_{pol,mm}(l,b)$ and $I_{mm}(l,b)$.

Obtaining a dust total emission map ($I_{mm}(l,b)$) is relatively easy
because dust total emission maps at FIR wavelengths are already available
{}from the {\it IRAS} and {\it COBE/DIRBE} observations.
Using the relation
\begin{equation}
     I_{mm}(l,b) = I_{100\mu m}(l,b)  ( 1mm/100\mu m )^{-\beta}, \label{eq:mm_100}
\end{equation}
where $1 \lesssim \beta \lesssim 2$,
one can easily obtain an emission map at {\it mm} wavelengths ($I_{mm}$) 
{}from the maps at 100 $\mu m$ or 240 $\mu m$.
However, more sophisticated model dust emission maps at {\it mm} wavelengths
already exist.
For example,
Finkbeiner et al. (1999) presented predicted full-sky maps of
microwave emission from the diffuse interstellar dust using
FIR emissions maps generated by Schlegel et al. (1998).
In fact, the model dust emission map we analyzed in the previous section
(\S\ref{sect:dust_em}) is one of the maps presented in 
Finkbeiner et al. (1999). 
Therefore, we can assume that the thermal dust emission map 
($I_{mm}(l,b)$) is already available.

Then how can we obtain 
a degree-of-polarization map at {\it mm} wavelengths ($P_{em,mm}(l,b)$)?
We can use measurements of starlight polarization at optical wavelengths
to get $P_{em,mm}(l,b)$. The basic idea is that
the degree of polarization by emission at mm ($P_{em,mm}$) is related to 
that at optical wavelengths ($P_{em,optical}$), which in turn is related to
the degree of polarization by absorption at optical wavelengths
($P_{abs,optical}$):
\be
  P_{abs,optical} \rightarrow P_{em,optical} \rightarrow P_{em,mm}.
\ee
 We describe the relations in detail below.

When the optical depth is small, we have the following relation (see, for
example, Hildebrand et al. 2000):
\begin{equation}
  P_{em,optical} \approx -P_{abs,optical}/\tau,   \label{eq:emabs}
\end{equation}
where $P_{em,optical}$ is the degree of polarization by emission and
$\tau$ is the optical depth (at optical wavelengths).
We obtain polarization by emission at {\it mm} wavelengths ($P_{em,mm}$) 
using the relation
\begin{equation}
  P_{em,mm}= P_{em,optical}
        \left[ \frac{ C_{max}-C_{min} }{ C_{max}+C_{min} } \right]_{mm}
      / \left[ \frac{ C_{max}-C_{min} }{ C_{max}+C_{min} } \right]_{optical},
  \label{eq_pem_mm}
\end{equation}
where $C$'s are cross sections (of grains as projected on the sky)
that depend on the geometrical shape (see, for example, the discussion in 
Hildebrand et al. 1999; see also Draine \& Lee 1984)
and dielectric function $\epsilon=\epsilon_1+i\epsilon_2$
(see Draine 1985) of 
grains.

For {\it mm} wavelengths, it is easy to calculate the ratio in Eq.~(\ref{eq_pem_mm})
because the wavelength $\lambda$ is much greater than the grain size $a$ 
(i.e.~$\lambda \gg 2\pi a$).
In this case, 
if grains are oblate spheroids with $a_1<a_2=a_3$
and 
short axes ($a_1$) of grains are perfectly aligned in the
plane of the sky, we have
\begin{equation}
    C_j=\frac{ 2\pi V }{ \lambda } 
        \frac{ \epsilon_2(\lambda ) }
        { \left( L_j \left[ \epsilon_1(\lambda )-1 \right] +1 \right)^2 
               + \left[ L_j \epsilon_2(\lambda ) \right]^2  },
\label{eq:c_j}
\end{equation}
where $L$ values are defined by
\bea
   L_1 &=& [(1+f^2)/f^2][1-(1/f)\arctan f], \nonumber \\
   L_2 &=& L_3=(1-L_1)/2, \nonumber \\
   f^2 &=& (a_2/a_1)^2-1
\eea
(see, for example, Hildebrand et al.~1999).

However, for optical wavelengths, the condition $\lambda \gg 2\pi a$ is {\it not}
always valid and, therefore, the expression in Eq.~(\ref{eq:c_j}) returns 
only approximate values.
For accurate evaluation of the cross sections, one should use numerical methods.
Fortunately, several numerical codes are publicly available for such 
calculations (for example, {\bf DDSCAT} package by Draine \& Flatau 1994, 2008; 
{\bf ampld.lp.f} by Mishchenko 2000).
We use {\bf ampld.lp.f} to calculate the ratio in Eq.~(\ref{eq_pem_mm}).
We assume that the grains are oblate spheroids, grain size is $0.1\mu$m, 
$\lambda_{optical}=0.5\mu$m, and $\lambda_{mm}=1000\mu$m.
Left panel of Fig.~\ref{fig:ratios} shows that the ratio is around $1.5$
when magnetic field is perpendicular to the line-of-sight. It also shows that
the ratio of 
$P_{em,mm}/P_{em,optical}$ is almost independent of the grain axis ratio.

In this subsection, we described a simple way to
obtain a polarized map at {\it mm} wavelengths.
However, actual implementation of the method can be more complicated
due to the following reasons.
First, we used an assumption that the grains that produce optical
absorption produce also microwave emission. But, this is not true
in general (see Whittet et al. 2008). 
Second, the expressions in Eqs.~(\ref{eq_pem_mm}) and (\ref{eq:c_j})
are valid when magnetic field direction is fixed and perpendicular 
to the line-of-sight and
all grains are perfectly aligned with the magnetic field.
If this is not the case, Eq.~(\ref{eq_pem_mm}) will become
$P_{em,mm} \propto P_{em,optical}$ with the constant of proportionality
that depends on magnetic field structure and the degree of
grain alignment. The effect of partial alignment is expected to be less
important\footnote{
 Theories (see Lazarian 2007 for a review) predict that
 grains starting with a particular size get aligned.
 If the grain size distribution varies from one place to another,
 this fraction of aligned grains will also vary.
 However, for the diffuse ISM, the grain distribution should not vary much
 (Weingartner \& Draine 2001).
 Therefore, the effect of partial alignment will be less significant.
}. Then, how serious is the effect of non-perpendicular magnetic field?
We perform a numerical calculation using {\bf ampld.lp.f} to evaluate
the effect.
We assume that the grains are oblate spheroids, grain size is $0.1\mu$m, 
$\lambda_{optical}=0.5\mu$m, and $\lambda_{mm}=1000\mu$m.
Right panel of Fig.~\ref{fig:ratios} shows that the polarization ratio
drops from  $\sim 1.5$ to $\sim 1.1$ when the angle (between magnetic field
and the plane of the sky) changes from $90^\circ$ to $\sim 5^\circ$.
Therefore, the effect is not very strong and can be potentially corrected for\footnote{
   We can make use of Right panel of Fig.~\ref{fig:ratios} reversely.
   In the future, when we can accurately measure
   polarized emission from thermal dust in FIR or $mm$ wavelengths,
   we can obtain the values of $[(C_{max}-C_{min})/(C_{max}+C_{min})]_{mm}$.
   This result combined with the values of 
   $[(C_{max}-C_{min})/(C_{max}+C_{min})]_{optical}$ 
   in optical wavelengths can be used to 
   find average angle between magnetic field and the plane of the sky.
   That is, when we know the ratio $[...]_{mm}/[...]_{optical}$,
   we can use Right panel of Fig.~\ref{fig:ratios} to find
   the angle between magnetic field and the plane of the sky.}.

\subsection{Angular spectrum of polarized emission from thermal dust}
After we have constructed a map of the polarized emission from thermal dust, 
we can obtain the angular spectrum.
However, if we are interested in the shape of the angular spectrum,
we do not need to construct the polarized thermal dust emission map.
We can get the shape of the the angular spectrum
directly from the starlight polarization map $P_{abs,optical}(l,b)$.

Eq.~(\ref{eq:IIP}) tells us that $I_{pol,mm}$ is given by  
$P_{em,mm}$ times $I_{mm}$. 
{}From Eqs. (\ref{eq:emabs}) and (\ref{eq_pem_mm}), we have
\bea
   I_{pol,mm}& = & P_{em,mm}~ I_{mm} \propto P_{em,optical}~ I_{mm}  \nonumber \\
   & \approx & (P_{abs,optical}/\tau)~ I_{mm} \nonumber  \\
   & \propto & P_{abs,optical}.  \label{eq:28}
\eea
Here we use the fact $\tau \propto I_{mm}$.
Note that the constant of proportionality does not affects the shape of the
angular spectrum if grain properties do not vary much in halo.
Therefore, as to the power spectrum $c_l$ of $I_{pol,mm}$, we can use
that of $P_{abs,optical}$:
\be
    C_l \mbox{~of $I_{pol,mm}$} \propto C_l \mbox{~of $P_{abs,optical}$}.
   \label{eq:29}
\ee
Once we know the angular spectrum of $P_{abs,optical}$,
we can estimate the angular spectrum of $I_{pol,mm}$.

Then, what is the measured angular spectrum of $P_{abs,optical}$(l,b)?
Fosalba et al.~(2002) 
obtained $C_l \sim l^{-1.5}$ for starlight polarization.
The stars used for the calculation are at different distances from
the observer and most of the stars are nearby stars.
The sampled stars are mostly in the Galactic disk.
CL02 reproduced the observed angular spectrum numerically
and showed
that the slope becomes shallower when only stars with a large fixed distance
are used for the calculation. 
Therefore, it is clear that distance, 
or dust column density, 
to the stars is an important factor
that determines the slope. 
We expect that, if we consider only the nearby stars with a fixed distance,
the slope will be steeper.
This means that, if we consider stars in the Galactic halo,
the slope will be steeper.

The method described above requires measurements of polarization from
many distant stars in the Galactic halo.
Unfortunately, the number of stars outside the Galactic disk that can be
used for this purpose
are no more than a few thousands (Heiles 2000; see also discussions in
Page et al. 2007; Dunkley et al. 2008).
When more observations are available in the future, accurate estimation
of $I_{pol,mm}(l,b)$ (and $C_l$ of $P_{abs,optical}$) will be possible.
We do not pursue this topic further in this paper.


\subsection{Model calculations for starlight polarization}
Then, what do we expect about $C_l$ of $P_{abs,optical}$ for the Galactic halo?
To deal with this problem we use numerical simulations again.
We first generate two sets of magnetic field on a two-dimensional
plane ($8192 \times 8192$ grid points), 
using Kolmogorov three-dimensional spectra\footnote{
   Consider a 3-dimensional magnetic field with a 3D spectrum
   $E_{3D}(k_x,k_y,k_z)$. 
   The spectrum of the magnetic field on a two-dimensional
   sub-plane (e.g.~$z=0$ plane) is 
   $E_{z=0 plane}(k_x,k_y)\propto 
     \int_{-\infty}^{\infty} dk_z~E_{3D}(k_x,k_y,k_z)$, which
   we use in our calculations to generate two sets of magnetic field
   on a two-dimensional plane.
}.
We consider 3 models:
\begin{enumerate}
\item {\bf Case 1}, Nearby stars in a homogeneous turbulent medium: 
       We generate three (i.e.~{\it x,y}, and {\it z}) 
       components of magnetic field on a two-dimensional
       plane ($8192 \times 8192$ grid points representing 
       $400 pc \times 400 pc$), 
       using the following Kolmogorov three-dimensional spectrum:
       $E_{3D}(k)\propto k^{-11/3}$ if $k>k_0$,
       where $k_0 \sim 1/100~pc$. 
     (The outer scale of turbulence is 100pc.)
      We assume the volume density of dust is homogeneous. 
      All stars are at a fixed distance of $100pc$ from the observer. 
\item {\bf Case 2}, Distant stars in a homogeneous turbulent medium:
       We generate three (i.e.~{\it x,y}, and {\it z}) 
       components of magnetic field on a two-dimensional
       plane ($8192 \times 8192$ grid points representing 
       $4~kpc \times 4~kpc$), 
       using the following Kolmogorov three-dimensional spectrum:
       $E_{3D}(k)\propto k^{-11/3}$ if $k>k_0$,
       where $k_0 \sim 1/100~pc$. 
      Other setups are the same as those of Case 1, 
      but the distance to the stars is $2kpc$.
\item {\bf Case 3}, Stars in a stratified medium: 
      We use the magnetic field generated in Case 1.
      The volume density of dust shows a $sech^2(z)$ decrease: 
      $\rho(r)=4\rho_0 /[ \exp(r/r_0)+\exp(-r/r_0) ]^2$.
      We assume spherical geometry and $r_0 = 100pc$. 
      The stars are at $r=200pc$ from the observer.
      The outer scale of turbulence is 100pc.
\end{enumerate}

We assume that dust grains are oblate spheroids.
In the presence of a magnetic field, some grains (especially large grains)
are aligned with the magnetic field (see Lazarian 2007 for a review). 
Therefore, cross sections parallel
to and perpendicular to the magnetic field are different.
We assume that parallel cross section is $\sim$30\% smaller than the 
perpendicular one.
We use the following equations to follow changes of Stokes parameters
along the path:
\bea
 I^{-1} dI/ds & = & -\delta + \Delta \sigma Q/I, \\
 d(Q/I)/ds & = & \Delta \sigma - \Delta \sigma (Q/I)^2, \\
 d(U/I)/ds & = & -\Delta \sigma (Q/I)(U/I)
\eea
(see Martin 1974 for original equations; see also Dolginov, Gnedin, \& Silantev 1996),
where $\delta = (\sigma_{1} + \sigma_{2})$,
$\Delta \sigma = (\sigma_{1} - \sigma_{2})$, and
\bea
2\sigma_{1} &=& \sigma_{\perp}, \\
2\sigma_{2} &=& \sigma_{\perp} -(\sigma_{\perp}-\sigma_{\|}) \cos\gamma
\eea
(Lee \& Draine 1985).
Here $\sigma_{\perp}$ and $\sigma_{\|}$ are the extinction coefficients 
and $\gamma$ is the angle
between the magnetic field and the plane of the sky. 
After we get the final values of Stokes parameters, we calculate
the degree of polarization ($\sqrt{Q^2+U^2}/I$) and, then, the second-order 
angular structure function of the degree of polarization.

\begin{figure*}[h!t]
\includegraphics[width=0.48\textwidth]{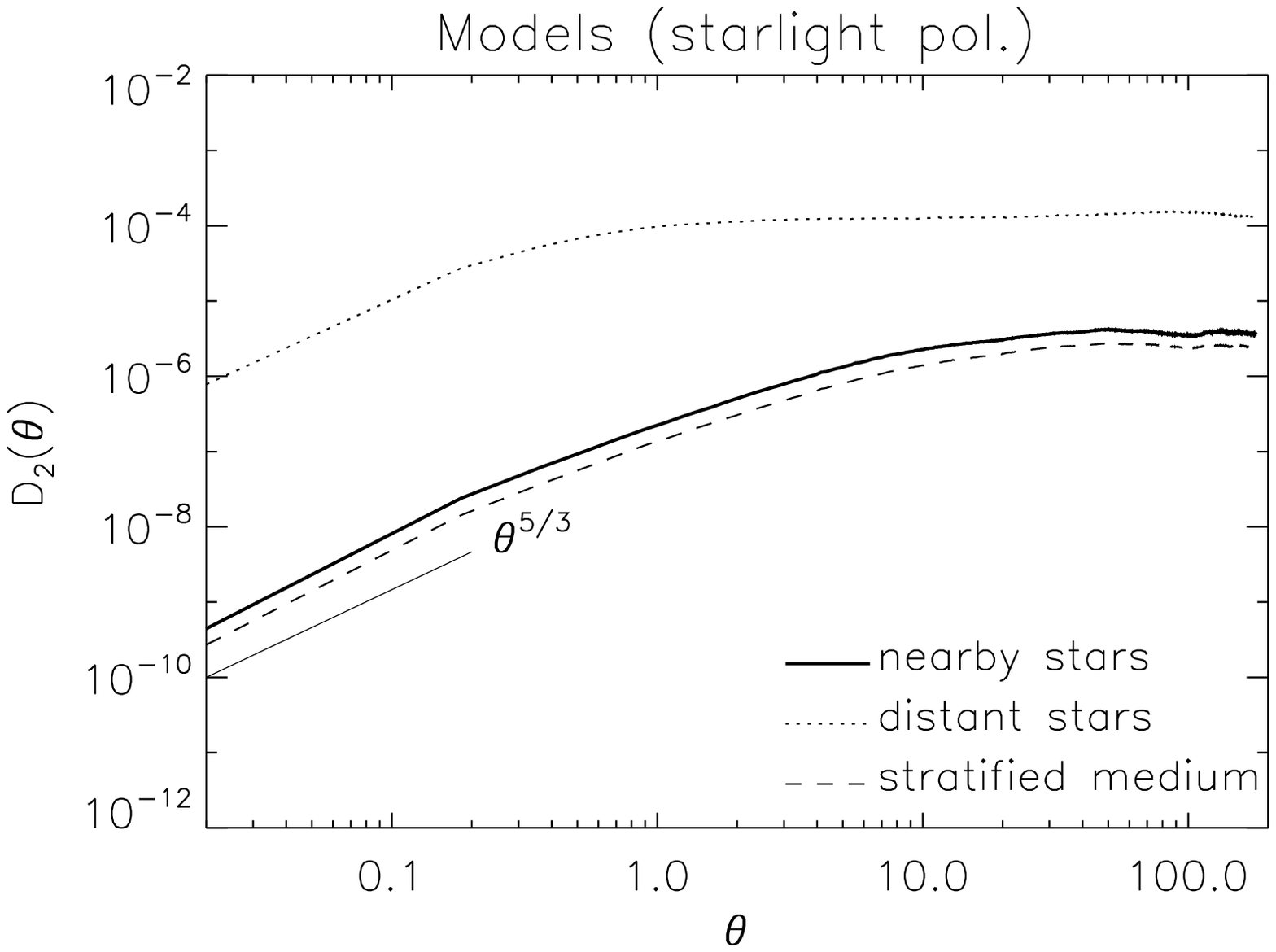}   
\includegraphics[width=0.48\textwidth]{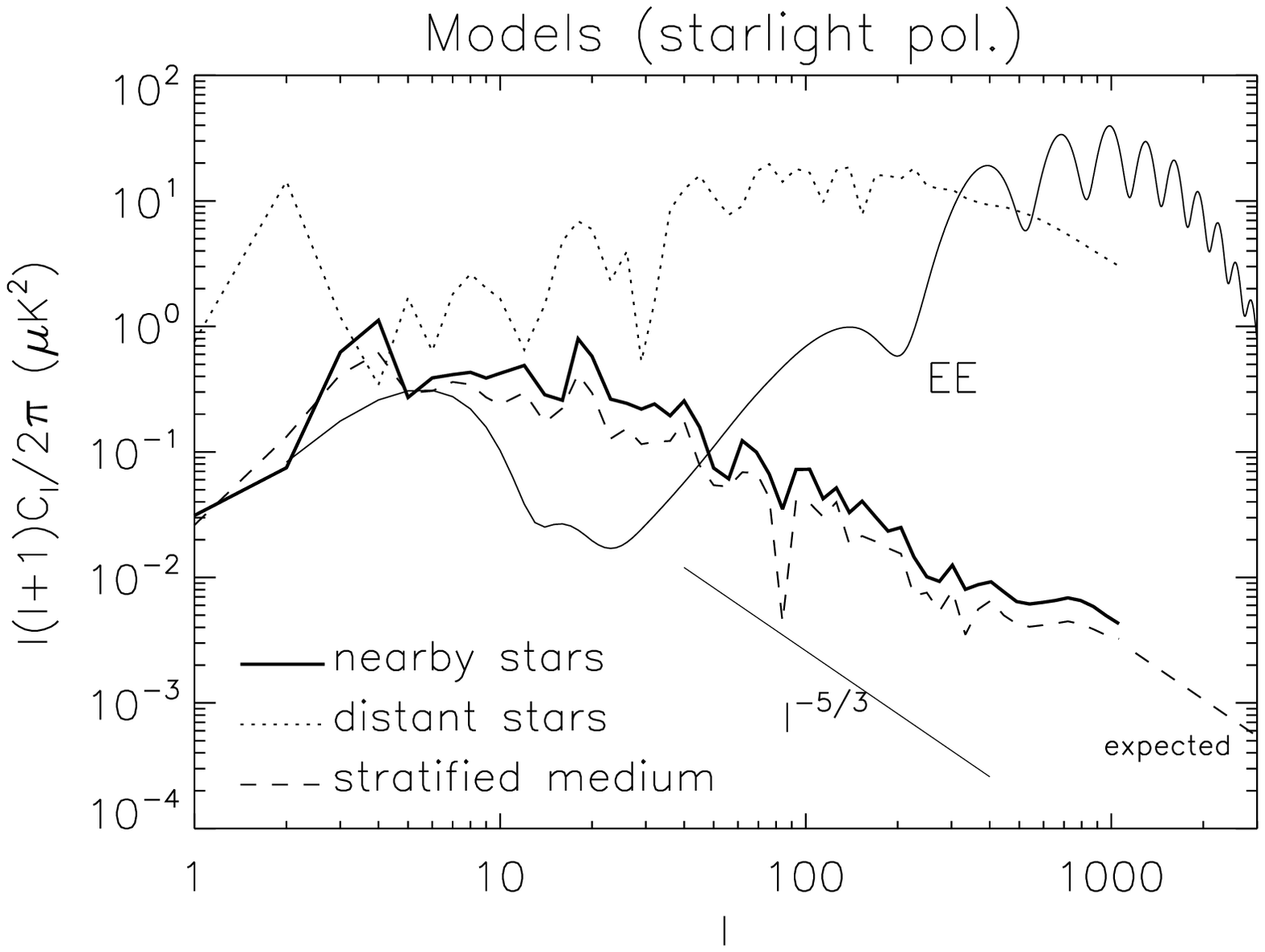}   
\caption{ The second-order angular structure function and angular power spectrum 
     for starlight polarization. Plots are for degree of polarization.
     {\it Left}: The second-order structure function for
     nearby stars (thick solid line; $d=100$pc; Case 1 in the text) 
     has a slope close to Kolmogorov one for $\theta \lesssim 0.2^\circ$ 
     and slightly
     shallower spectrum for $0.2^\circ \lesssim \theta \lesssim 10^\circ$.
     The case for density stratification 
    (dashed line; scale height $= 100$pc; Case 3 in the text)
     shows also a similar slope.
     However, the case for far-away stars 
    (dotted line; $d=1$kpc; Case 2 in the text)
     shows a substantially shallower slope for $\theta \gtrsim 0.1^\circ$.
     {\it Right}: Nearby stars (thick solid line) or 
        stars in stratified medium (dashed line) 
        show a slope flatter than
        the Kolmogorov when $l\lesssim 1000$. 
        We do not show spectra for $l>1000$, because data are too noisy.
        The second-order structure function on the left panel implies that
        $l(l+1)c_l \propto l^{-5/3}$ for $l>1000$.
        The spectra are normalized by the condition 
        $\sum_{2}^{10} l(l+1)c_l/2\pi = 3 (\mu K^2)$
        for the dashed line, which we believe is the most realistic model 
        among the 3 toy models.
        We assume that the observed frequency is $94 GHz$.
     Note that $\theta=0.1^\circ$ corresponds to 
      $l\sim 180^\circ/\theta^\circ \sim 1800$.
     We also show the polarized CMB `EE' spectrum.
 }
\label{fig:starpol}
\end{figure*}

We show the result in Fig.~\ref{fig:starpol}.
When the stars are nearby (Case 1), the spectrum is consistent with the Kolmogorov
spectrum for small $\theta$.
The result for the stratified medium (Case 3) also shows a spectrum compatible with
the Kolmogorov one for small $\theta$.
When stars are far away (Case 2), the qualitative behavior is similar.
However, if we measure average slope between $\theta = 0.2^\circ$ and $20^\circ$, 
the result is different:
the slope for Case 2 is substantially shallower.
Note that $\theta = 0.2^\circ$ and $20^\circ$
correspond to $l=1000$ and $10$, respectively.
This means that,
 when we have either distant stars  or mixture of distant and nearby stars,
the angular spectrum will be shallower than the Kolmogorov one.
Therefore, it is not surprising that Fosalba et al.~(2002) obtained
a shallow spectrum of $\sim C_l\propto l^{-1.5}$ 
for stars mostly in the Galactic disk.
Flattening of spectrum (i.e. $C_l \propto l^{-\alpha}$ with
$\alpha \approx 1.3$ $\sim$ $1.4$) for polarized FIR dust thermal emission is also
observed in
Prunet et al.~(1998; see also Prunet \& Lazarian 1999).

\subsection{Comparison with the CMB polarization}
On the right panel of Fig.~\ref{fig:starpol}, we plot angular power spectrum
of star light polarization.
As we mentioned earlier, the angular spectrum of the degree of starlight polarization
should be similar to that of polarized thermal dust emission 
(see Eqs.~[\ref{eq:28}] and [\ref{eq:29}]).

To obtain angular spectra, we use a Gauss-Legendre quadrature integration
method as described in Szapudi et al. (2001).
To be specific, we first generate magnetic fields from the 3 models we considered 
in the previous subsection. 
Then, we calculate angular correlation functions, $K(\cos\theta)$.
Finally, we obtain the angular spectra using Eq.~(\ref{eq:k2cl}).
Since $C_l$ obtained in this way is very noisy, we plot $C_l$ averaged over the multipole range 
$(l/1.09, 1.09l)$.
We do not show $C_l$ for $l>1000$ because it is too noisy even with the averaging process.
We normalize the spectra using the condition
$\sum_{2}^{10} l(l+1)C_l/2\pi = 3 (\mu K^2)$.
This normalization is based on the values given in Page et al.~(2007; their Eq.~[25]).
We assume that the observed band is W-band ($\nu = 94 GHz$).

The plot shows that the slopes for nearby stars (thick solid line) 
and stars in the stratified medium are shallower than that of 
the Kolmogorov spectrum
for $l<1000$. 
This result is consistent with that obtained with the angular structure function.
Note that the case of distant stars has much flatter spectrum for $l<1000$.
We believe that our toy model for the stratified medium (Case 3) 
better represents the actual situation for polarized emission from thermal 
dust in the Galactic halo. 
Therefore, 
we expect that the polarized thermal emission from thermal dust
in high-latitude Galactic halo has a spectrum slightly shallower than the Kolmogorov spectrum
for $l<1000$.
Our calculation do not tell us about the slopes for $l>1000$.
However, judging from the behavior of the angular structure function 
for $\theta \lesssim 0.1^\circ$,
we expect that $C_l \propto l^{-11/3}$ for $l>1000$
(see the straight dashed line for $l>1000$ on the right panel of Fig.~\ref{fig:starpol}).

The Figure shows that the EE spectrum dominates polarized thermal dust emission from high-latitude
Galactic halo for
$l\gtrsim 100$. The EE spectrum is expected to be sub-dominant when $l>5000$.

\section{Underlying spectrum: Additional information for removing
foregrounds}

Removal of Galactic foregrounds has always been a big concern for CMB
studies. The challenge is only going to increase substantially now, when CMB
polarization studies are attempted.

The knowledge that the foreground are not an arbitrary noise, but have well
defined statistical properties in terms of their {\it spatial power spectra}
is an important additional information that can be utilized to evaluate and
eventually eliminate the foreground contribution. 

Utilizing the information about underlying turbulence power spectrum is not
straightforward, however. Our study shows that the observed power spectrum
may depend on geometry of the emitting volume. Therefore, the detailed
modeling of the foreground fluctuations should involve accounting for the
geometry of the emitting volume.

The latter point stresses the synergy of the Galactic foreground and CMB
studies. Indeed, our fitting of the power spectra 
in Fig.~\ref{fig:model} shows that on the
basis of its variations we may distinguish between different models of the
emitting turbulent volume. As soon as this achieved, one can {\it predict},
for instance
the level of fluctuations that are expected from the foreground at the
scales smaller than those studied. A simplification that is expected at
higher resolutions that are currently available, is that at sufficiently
small scales the statistics should get independent of the large-scale
distributions of the emitting matter.

Consider an example of utilization of this approach. 
As we know, {\it Planck} will
sample the polarized foregrounds with the resolution of 
up to $l\sim 2000$ ($\delta \theta \sim 5^\prime$). For {\it Planck} high
frequency coverage should help to remove the foregrounds with high
accuracy\footnote{Even in the case of {\it Planck} checking the consistency of the
foreground with the model of emission from a turbulent medium with a given
geometry may be a useful test, however.}. 
A higher resolution will be
available, for instance, with balloon-borne missions, which in many
cases will not have as many frequency channels as {\it Planck} does. However, our
present work suggests that the contribution of the foregrounds to the
spectrum measured by these high-resolution missions can be evaluated on the
basis of our knowledge of the underlying {\it spatial power spectrum}. 

Consider, for instance, Fig.~\ref{fig:starpol}. If {\it Planck} measures the spatial spectrum up to
$l=2000$, then for a higher resolution balloon mission one can evaluate the
level of foreground contamination by extrapolating the expected foreground
spectrum. 

While most of the paper is directly related to making use of the knowledge
of the underlying spectra and/or 
two points correlations in order to deal with the
foregrounds, the part dealing with higher-order statistics is not directly
related to the foreground removal. Nevertheless, our analysis shows that
high-order correlations can also help in understanding of the properties of
foregrounds.

\section{High-order statistics}  \label{sect:h_o}
High-order structure functions 
are used for the study of intermittency, which refers to the non-uniform distribution of structures.
The structure functions of order $p$ for an observable $I$ is defined by
\be
   S_p(r)=< |I(x)-I(x+r)|^p>,
\ee
where the angled brakets denote average over position $x$.
For an observable defined in the plane of the sky, 
the angular structure function of order $p$ is
\be
   D_p(\theta)=< |I({\bf e}_1)-I({\bf e}_2)|^p>,
\ee
where ${\bf e}_1$ and ${\bf e}_2$ are unit vectors along the lines of
sight and $\theta$ is the angle between ${\bf e}_1$ and ${\bf e}_2$.

Traditionally, researchers use
high-order structure functions of velocity to probe 
dissipation structures of turbulence.
In fully developed hydrodynamic turbulence, the (longitudinal)
velocity structure functions
$S_p=< ( [ {\bf v}({\bf x}+ {\bf r}) -
      {\bf v}({\bf x})]\cdot \hat{\bf r} )^p>
\equiv < \delta v_L^p({\bf r}) >$ are
expected to scale as $r^{\zeta_p}$.
One of the key issues in this field is the functional form of the
scaling exponents $\zeta_p$.
There are several models for $\zeta_p$.
Roughly speaking, the dimensionality 
of the energy dissipation structures plays an important role.

Assuming 1-dimensional worm-like 
dissipation structures, She \& Leveque (1994) 
proposed a scaling relation
\be
  \zeta_p^{SL}=p/9+2[1-(2/3)^{p/3}]
\ee
for incompressible hydrodynamic turbulence.
On the other hand, assuming 2-dimensional sheet-like dissipation structures,
M\"uller \& Biskamp (2000) proposed the relation
\begin{equation}
\zeta_p^{MB}=p/9+1-(1/3)^{p/3}
\end{equation}
for incompressible magneto-hydrodynamic turbulence\footnote{
   Boldyrev (2002) obtained the same scaling relation for 
   highly supersonic turbulence. However, since it is unlikely that
   turbulence in the Galactic halo is highly supersonic, we refer the
   scaling relation to 
   the ``Muller-Biskamp'' scaling.
}.

Recently, high-order structure functions of molecular line intensities
have been also employed (Padoan et al. 2003; 
Gustafsson et al. 2006).
In optically thin case, the molecular line intensities are proportional
to the column density.
Kowal, Lazarian, \& Beresnyak (2007) studied scaling of
density fluctuations in MHD turbulence.
Their numerical results show that behavior of the scaling exponents 
for column density depends on sonic Mach number of turbulence.

The Haslam map shows a reasonable agreement with the Muller-Biskamp 
MHD model.
In some sense this is natural because the synchrotron emission
arises from MHD turbulence in the Galactic halo and roughly reflects 
the column density of cosmic-ray electrons.
However, it is not clear why the column density shows a similar scaling
 as velocity.
  Note that Padoan et al. (2003) also obtained 
a similar result using $^{13}CO$ emission from Perseus and Taurus.
However, the physical origin may be different: their result may 
reflect existence of highly supersonic turbulence in those clouds.

Unlike the Haslam map, the dust map does not show agreement with
existing models. The scaling exponents do not show strong dependence
on the order $p$. 
Left and middle panels of Fig.~\ref{fig_10} 
   show that the slope is around 1 for high-order structure functions.
This kind of behavior is expected
when discrete structures dominate the map.
However, it is not clear what kinds of discrete structures dominate. 
Either thin filamentary structures or point sources 
could explain the behavior of the structure functions\footnote{
   Consider a circular uniform cloud with
   radius $\Delta$ and intensity $I$ centered
   at the origin.
   (For simplicity, let us consider the 2D Cartesian coordinate system.)
   Then we can show that $D_n(r) \propto I^n (2\pi r)\Delta$
   for $r\gg \Delta$. When $r\sim \Delta$, $r$-dependence will be weaker.
   We can show that structure functions for a filament also show 
   a similar $r^1$-dependence,
   because a filament can be viewed as a chain of
   circular clouds (or a chain of square-like clouds):
    $D_n(r) \propto I^n (2\pi r)L_f \Delta$, where $L_f$ is the length
   of the filament.
   Therefore, both filaments and cloud-like sources exhibit $r^1$-scaling.
   The high-order structure functions for the dust emission map
   show $\sim r^1$ power-law scaling
   for $\theta \gtrsim 1^\circ$ (left panel of Fig.~\ref{fig_10}).
   Thus the dominant discrete sources are either thin filaments
   or point sources.
  We expect the typical size or width of the dominant 
   structures is $\lesssim 1^\circ$.
}.
Since the Haslam map and the dust map sample different types of 
the ISM, it is not
so surprising that they show different scaling behaviors.

For the Galactic disk, high-order structure functions
of both maps show nearly flat structure functions.
The structure functions show a nearly flat behavior 
even for $\theta \sim 1^\circ$, which means that
the typical size or width of the dominant structures is $\sim 1^\circ$
or smaller.
Again, this kind of behavior is expected
when thin discrete filamentary structures or point-like sources 
dominate in the Galactic disk. 
Note that, since the region of the sky we considered is only a thin 
stripe ($-2^\circ < b <2^\circ$) along the Galactic disk, 
the calculation domain becomes 1D-like\footnote{
   In 1D space, a ``point-source'' produces
   completely flat structure functions.
} and
it is possible that
thin filaments and point sources of size $\lesssim 1^\circ$ can exhibit
a slope flatter than $r^1$.

\begin{figure*}[h!t]
\includegraphics[width=0.32\textwidth]{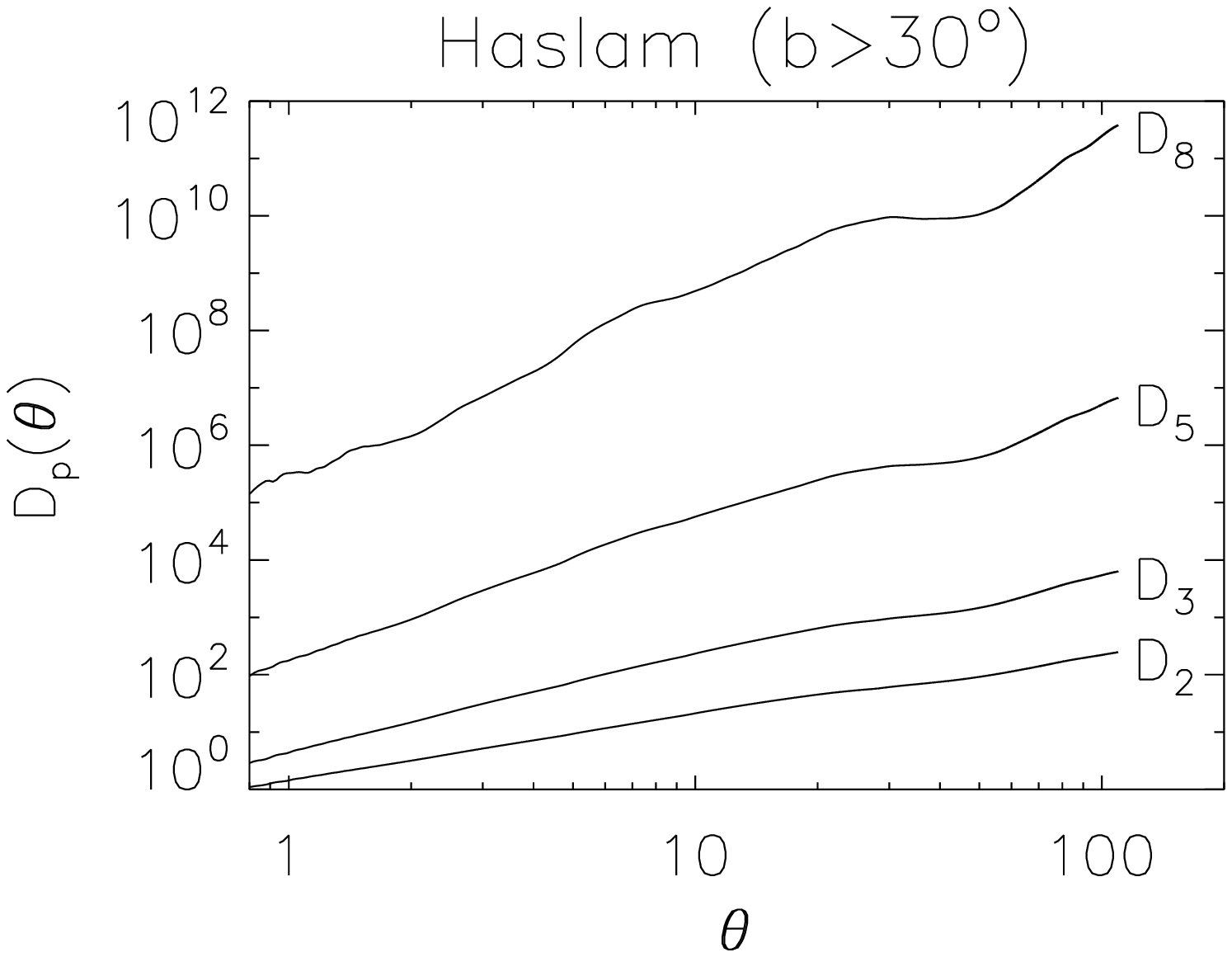}   
\includegraphics[width=0.32\textwidth]{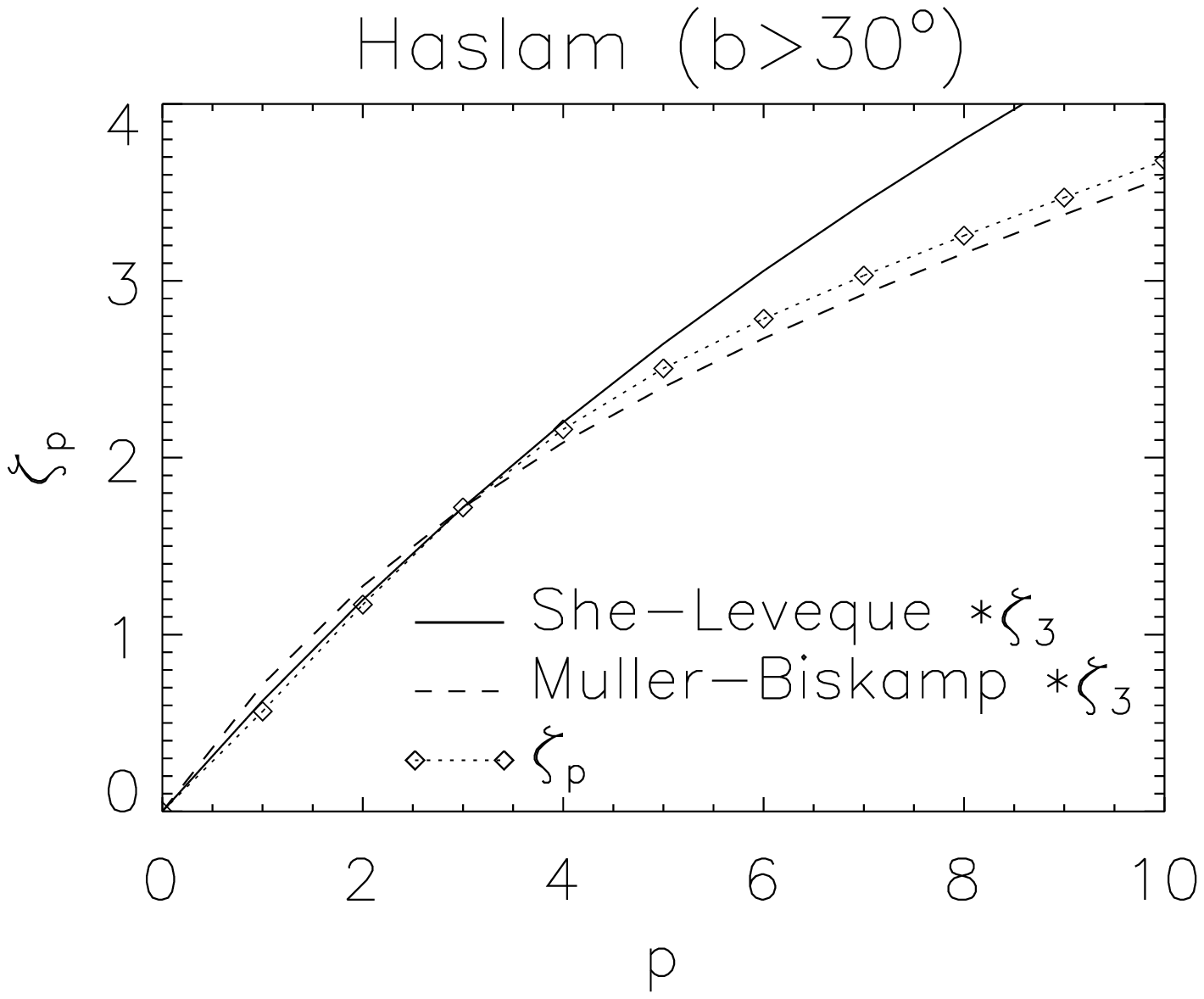}  
\includegraphics[width=0.32\textwidth]{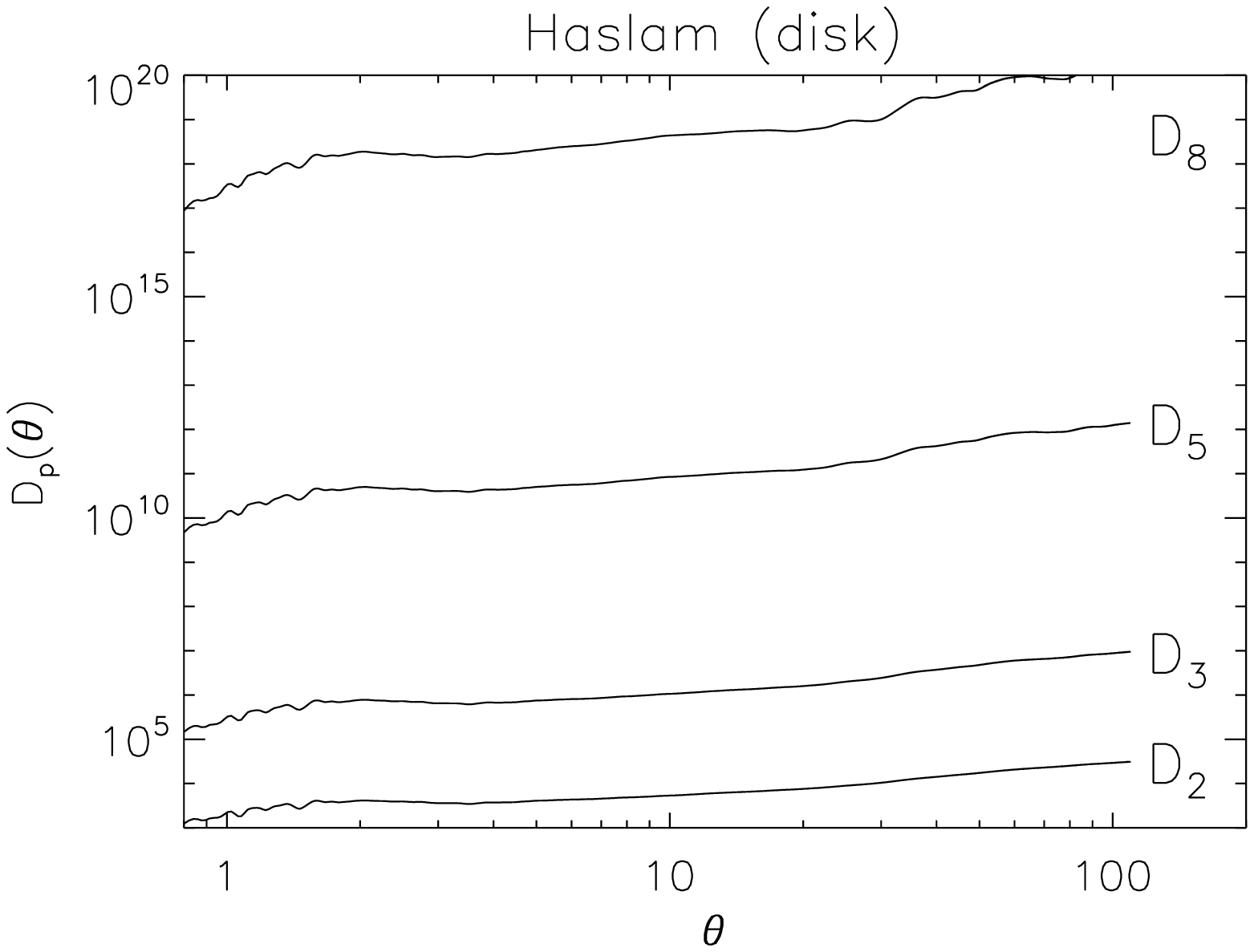}   
\caption{ 
   High order statistics of the Haslam map.
 {\it Left}: The second, third, 5th, and 8th-order structure functions
     for $b>30^{\circ}$.
 {\it Middle}: The scaling exponents for $b>30^{\circ}$
      seem to follow the Muller-Biskamp  
      MHD scaling.
      Note that we equate the scaling exponent of the observed third-order
      structure function and that of the Muller-Biskamp model.
      We measure the slope between $\theta=1^{\circ}$ and $10^{\circ}$.
 {\it Right}: The second, third, 5th, and 8th-order structure functions
     for the Galactic disk ($ -2^\circ \le b \le 2\circ$).
     They are all flat.
}
\label{fig_10}
\end{figure*}

\begin{figure*}[h!t]
\includegraphics[width=0.32\textwidth]{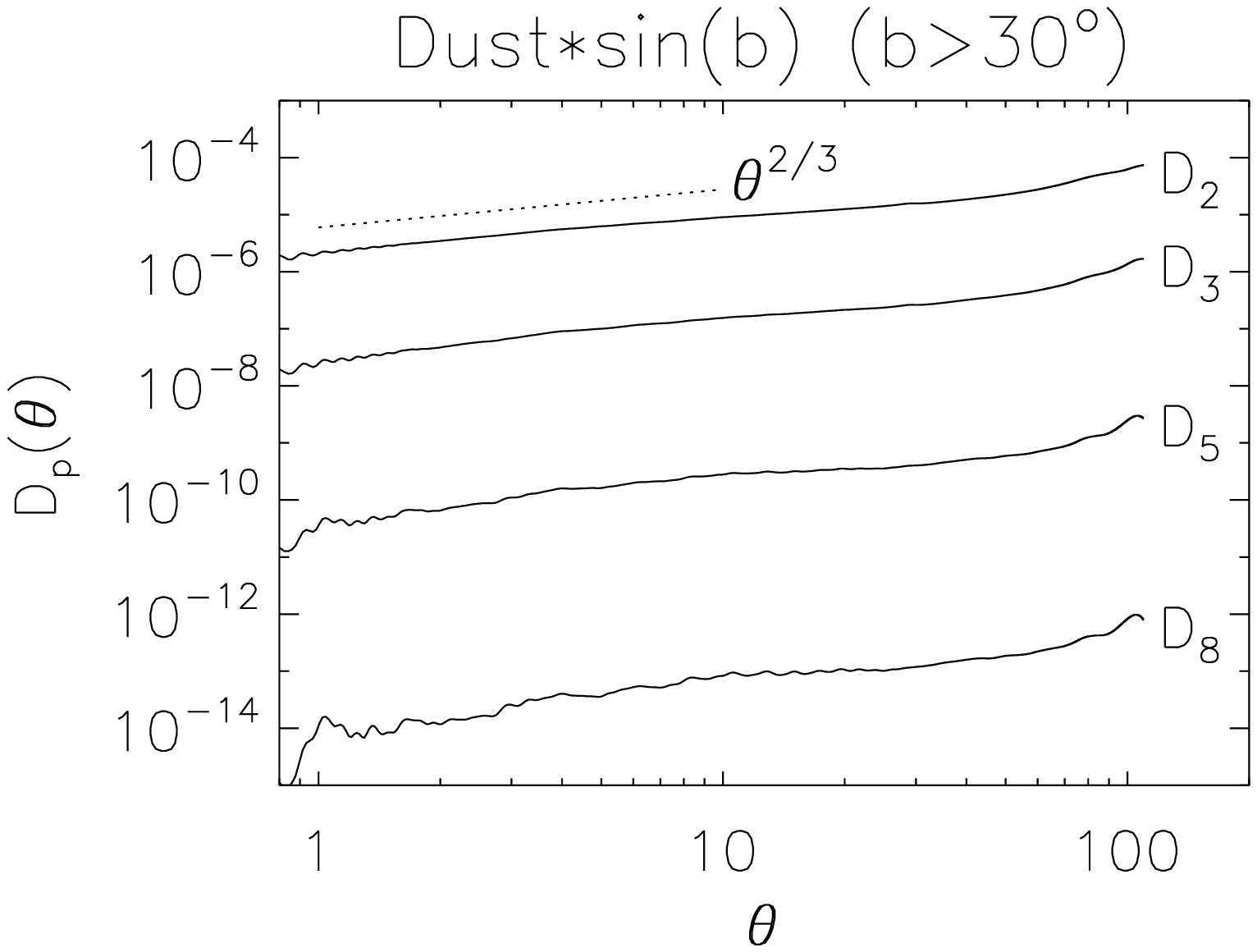}   
\includegraphics[width=0.32\textwidth]{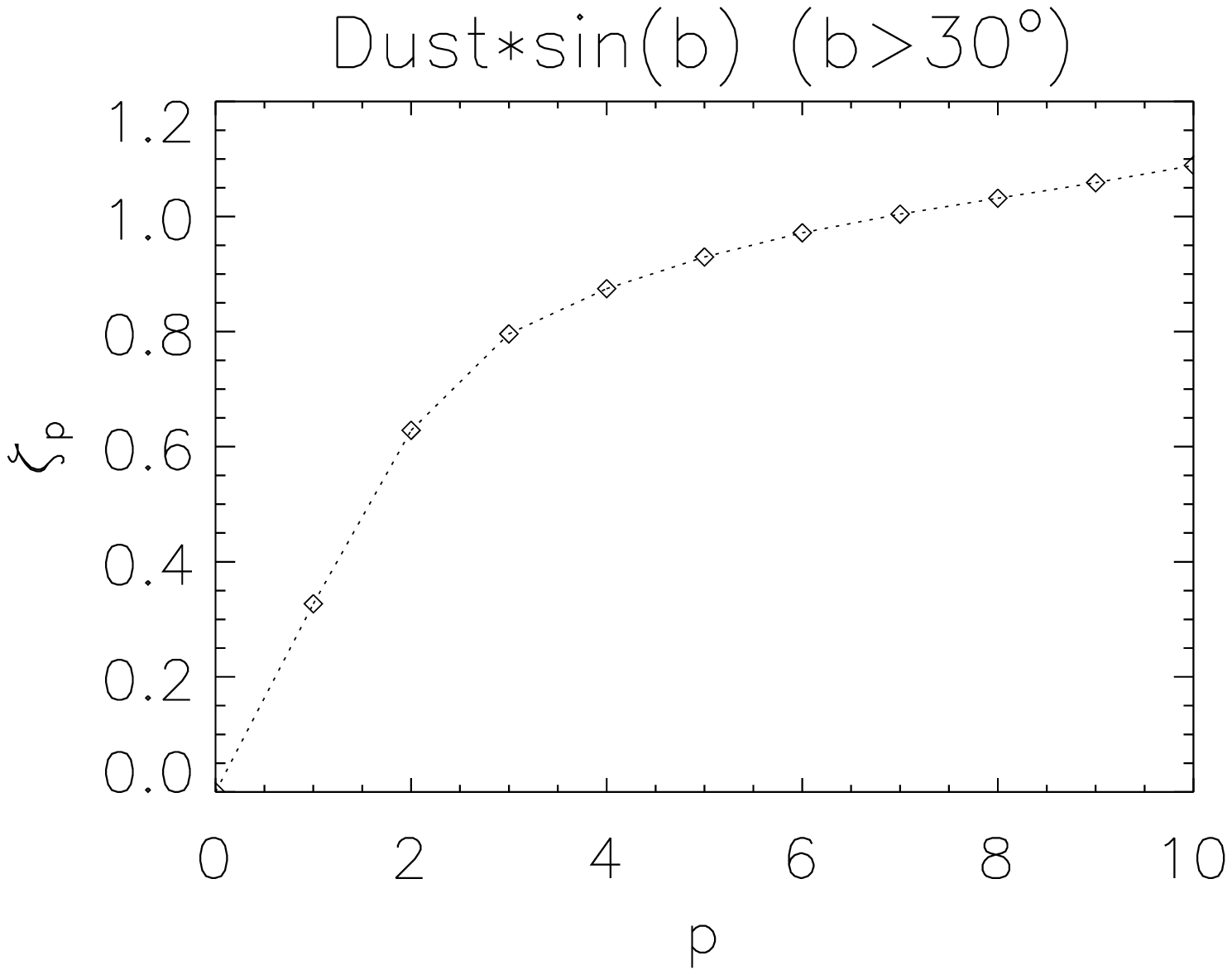}   
\includegraphics[width=0.32\textwidth]{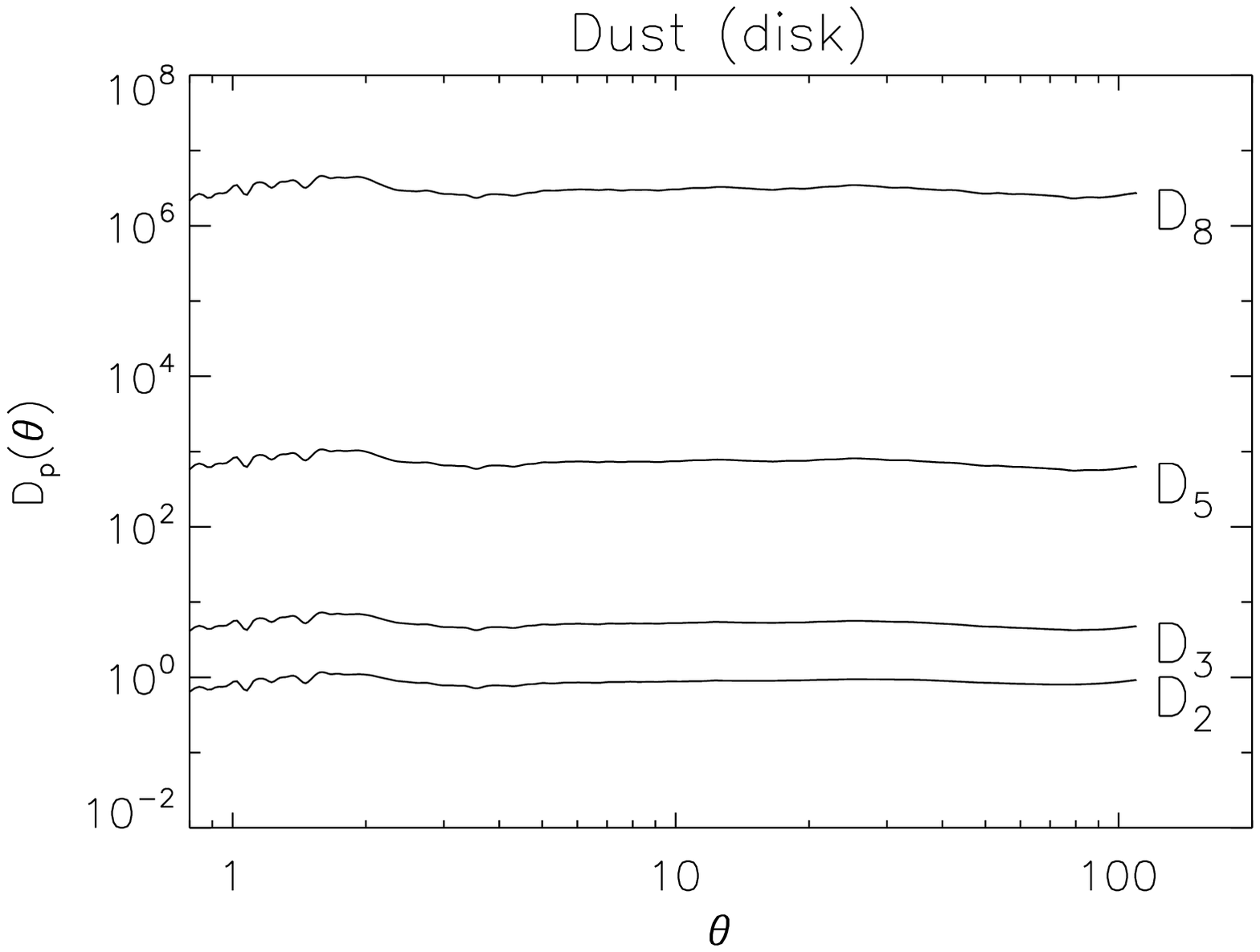}   
\caption{ 
   High order statistics of the model dust emission map. 
   We use dust emission intensity times $\sin b$ for all calculations.
 {\it Left}: The second, third, 5th, and 8th-order structure functions
     for $b>30^{\circ}$. 
 {\it Middle}: The scaling exponents for $b>30^{\circ}$
      does not show signatures of turbulence.
      We measure the slope between $\theta=1^{\circ}$ and $10^{\circ}$.
 {\it Right}: The second, third, 5th, and 8th-order structure functions
     for the Galactic disk ($ -2^\circ \le b \le 2\circ$).
     They are all flat.
}
\label{fig_10}
\end{figure*}

\section{Discussion and Summary}
In this paper, we have discussed the relation between 3D spatial spectrum and
the observed angular spectrum (or the second-order angular structure function) 
of the CMB foreground emissions. 
We have focused on synchrotron total intensity and
polarized thermal dust emission.
Our current study, as well as earlier studies (Chepurnov 1999;
CL02), predicts that $C_l$ will reveal true turbulence spectrum on small
angular scales. Then, on what scales will we see the turbulence spectrum?
Our model calculations that take into account stratification effects
imply that
\begin{enumerate}
\item $\theta <$ a few times $0.1^\circ$ (or $l>$ a few times $100$) for synchrotron emission (see Fig.~\ref{fig:model}), and
\item $\theta \lesssim 0.1^\circ$ (or $l\gtrsim 1000$) for polarized emission
{}from thermal dust (see Fig.~\ref{fig:starpol}).
\end{enumerate}
On larger angular scales, spectra are expected to be shallower.

Then, how is the turbulence spectrum related to the angular spectrum
(or the second-order angular structure function) on small scales?
When the 3D spatial spectrum has a power-law spectrum, $k^{-m}$, the
observed angular spectrum and the second-order angular structure function will be
\bea
   C_l & \propto & l^{-m}   \mbox{~~~ and} \\
   D_2(\theta) & \propto & \theta^{m-2} 
   \mbox{~~~~ if $\theta \ll L/d_{max}$},
\eea
where $L$ is the outer scale of turbulence and $d_{max}$ is the distance to the
farthest eddy in case of homogeneous turbulence.
We can interpret $d_{max} \sim z_0$ if there is a stratification with 
scale height $z_0$ (see Fig.~\ref{fig:model}).
When the angular separation is large, we have a universal scaling
for the angular correlation function:
\be
   K(\theta) \propto  \frac{ \pi -\theta }{ \sin\theta } \sim \frac{1}{\theta}
   \mbox{ ~~~if $\theta \gtrsim L/d_{max}$}
\ee
In this limit, the structure function is roughly constant and the angular spectrum is
roughly proportional to $l^{-1}$.

In this paper, we have analyzed Haslam 408Mhz map and a
model dust emission map and compared the results with model calculations. 
We have found that 
\begin{enumerate}
\item The Haslam map for high galactic latitude ($b>30^{\circ}$) can be explained by 
      MHD turbulence in the Galactic halo. The measured second-order angular
      structure function is proportional to $\theta^{1.2}$, which corresponds to
      an angular spectrum of $l^{-3.2}$.
      The high-order statistics for high galactic latitude ($b>30^{\circ}$) is
       consistent with that of incompressible 
       magnetohydrodynamic turbulence.
       Our model calculations show that a two-component model 
      (see \S\ref{sect:3models} and Fig.~\ref{fig:model})
       can naturally explain the observed angular spectrum.
       The one-component model can also explain the observed slope.
       But, the slope of the spectrum shows a more abrupt change
       near $l\sim 30$.

\item The model dust emission map may not have anything to do with turbulence on large angular scales.
      That is, we do not find signatures of turbulence in the map.

\item Both maps show
       flat high-order structure functions for the Galactic plane.
       This kind of behavior is expected
       when discrete structures dominate the map.
\end{enumerate}

We have described how we can obtain angular spectrum of polarized emission from
    thermal dust in high galactic latitude regions.
Our model calculations show that starlight polarization arising from dust in high galactic latitude regions
will have a Kolmogorov spectrum, $C_l\propto l^{-11/3}$, 
for $l\gtrsim 1000$ and
a shallower spectrum for $l\lesssim 1000$ (Fig.\ref{fig:starpol}).
We expect that polarized emission from the same dust also 
has a similar angular spectrum.
That is, we expect that the angular spectrum of polarized emission from thermal dust
is close to a Kolmogorov one for $l\gtrsim 1000$.

\acknowledgements
J.C.'s work was supported by the Korea Research Foundation grant
funded by the Korean Government (KRF-2006-331-C00136). 
A. Lazarian acknowledges the support by the NSF grants AST AST 0507164, as well as by the
NSF Center for Magnetic Self-Organization in Laboratory and Astrophysical 
Plasmas. 
The works of Jungyeon Cho was also supported by KICOS through
the grant K20702020016-07E0200-01610 provided by MOST.
We acknowledge the use of the Legacy Archive for Microwave Background Data Analysis (LAMBDA). Support for LAMBDA is provided by the NASA Office of Space Science.
We thank Alexei Chepurnov and John Everett for valuable discussions.

\appendix
\section{A. The  second-order angular structure function in the small angle limit}
The angular correlation $K(\theta)$ is given by the integral
\be
   K(\theta)=\int dl_1 \int dl_2 ~{\cal K}(|{\bf l}_1 - {\bf l}_2|),
\ee
where ${\cal K}(r)$ is the 3 dimensional spatial correlation.
Suppose that ${\bf l}_1$ is along x-axis, ${\bf l}_1=(l_1,0)$,  and 
${\bf l}_2 = l_2 (\cos \theta, \sin \theta)$.
Then, the correlation in the limit of small $\theta$ is given by
\bea
   K(\theta) & = &\int dl_1 \int dl_2 
              ~{\cal K}\left( \sqrt{ (l_1-l_2 \cos\theta)^2 + l_2^2 \sin^2\theta } 
      \right) \\
      & = & \int dl_1 \int dl_2 
              ~{\cal K} \left( \sqrt{ (l_1^2-2l_1 l_2 \cos\theta + l_2^2 } \right) \\
      & \approx & \int dl_1 \int dl_2
              ~{\cal K} \left( \sqrt{ (l_1- l_2)^2 + l_1 l_2\theta^2 } \right).
\eea
Suppose that the spatial correlation follows a power law: ${\cal K}(r) \propto \mbox{~const}-r^p$ 
for $r < L$, where $L$ is the outer-scale of turbulence.
For Kolmogorov turbulence, $p=2/3$.
Then the derivative of $K(\theta)$ is given by
\bea
  \frac{ K(\theta) }{ d\theta } & \propto &
      -\int dl_1 \int dl_2
              ~\left[  (l_1- l_2)^2 + l_1 l_2\theta^2  \right]^{p/2-1} (2l_1l_2\theta) \\
   & -\propto &
      \int du \int dw
              ~\left[  w^2 + (u^2-w^2) \theta^2 /4 \right]^{p/2-1} (u^2-w^2)\theta /2 ,
\eea
where $u=l_1+l_2$ and $w=l_1-l_2$.
If $p \leq 1$, the integration diverges as $\theta$ goes to zero.\footnote{
    When $p=1$, the spatial correlation becomes ${\cal K}(r)\propto C-r$, where
    $C$ is a constant.
    The corresponding 3D spectrum is $E(k) \propto k^{-4}$.
    When the slope of the turbulence spectrum is 
    steeper than $k^{-4}$,
    the correlation function has the form 
    ${\cal K}(r) \propto {\cal K}_0-r^{1}$ regardless
    of the turbulence slope.
    On the other hand, when the three-dimensional spectrum of turbulence 
    is shallower than $k^{-4}$, we have
    ${\cal K}(r) \propto {\cal K}_0-r^{m-3}$, where ${\cal K}_0\sim L^{m-3}$
    is a constant.
    Therefore, the condition of $p\leq 1$ is generally satisfied in turbulent
    medium.
}
Therefore, when $p \leq 1$, it suffices to perform the integration in the vicinity
of $l_1=l_2$ or $w=0$. Then we have
\bea
 \frac{ K(\theta) }{ d\theta } & \propto &
     -\int du \int dw
             ~\left[  w^2 + u^2 \theta^2 /4 \right]^{p/2-1} u^2\theta /2 \\
    & \approx & -\int du ~(u \theta/2)^{p-1} u^2 \theta /2
     \propto  -\theta^p,
\eea
where we use $\int_{-\infty}^{+\infty} dw/(w^2+A^2)^n = 
    A^{1-2n}\int_{-\pi/2}^{+\pi/2} d\theta ~\sec^{2-2n}\theta $.
Therefore, for small $\theta$ we have
\be
  K(\theta) \propto C_1 - C_2\theta^{p+1},
\ee
where $C_1$ and $C_2$ are constants.
Comparing this equation with 
\be
  K(\theta)=\mbox{$C_3$}-C_4 D_2(\theta),
\ee
we get
\be
   D_2(\theta) \propto \theta^{p+1}.
\ee

Analytic expressions for the relation between the angular structure
function ($D_2$) and the spatial 1D spectrum (E(k); in case of Kolmogorov, 
$E(k)\propto k^{-5/3}$) 
can be found in the literature.
For example, Lazarian (1995a; see also Lazarian \& Shutenkov 1990) 
derived the following expression:
\be
 E(k) \propto k \int_0^{{\cal L}/R} d\eta~ \frac{ d }{ d\eta } 
    \left( Q(\eta)\eta \right) J_1(kR\eta) + {\cal K}_5,
\ee
where ${\cal L}$ can be regarded as the outer scale of turbulence,
$R$ is the size of the system,
$Q(\eta)\sim D_2^\prime(\eta)\eta$, $\eta=\sin\theta$, $J_1(x)$
is the Bessel function of the first order, and ${\cal K}_5$
is a small correction term.

\section{B. Spatial spectrum of emissivity}
The synchrotron emissivity is proportional to $\sim n(e)B^\gamma \propto B^2$, where
$n(e)$ is the high-energy electron number density.
Suppose that magnetic field is roughly a Gaussian random variable.
This may not be exactly true, but should be a good approximation.
When a Gaussian random variable $B({\bf r})$\footnote{
    For simplicity, we assume $B$ is a scalar.
}
follows a Kolmogorov spectrum
\be
  E_{B,3D}\equiv |\tilde{B}(k)|^2 \propto \left\{ \begin{array}{ll} 
                              0                        & \mbox{if $k\le k_0$} \\
                              (k/k_0)^{-11/3}          & \mbox{if $k\ge k_0$,}
                      \end{array}
              \right. \label{B_E_3D}
\ee
we can show that the 3D spectrum of $B^2(r)$ follows Eq.~\ref{E_3D} (see, for example,
Chepurnov 1999).
The correlation of $B^2(r)$ and 3D energy spectrum of $B^2(r)$ are related by
\bea
     {\cal K}_{B^2}({\bf r})=< B^2({\bf x})B^2({\bf x}+{\bf r}) >_x \propto 
          \int  E_{B^2, 3D}({\bf k}) e^{i{\bf k}\cdot {\bf r}} d^3{\bf k}, \\
     E_{B^2, 3D}({\bf k})\equiv |\tilde{B^2}(k)|^2 \propto 
           \int {\cal K}_{B^2}({\bf r}) 
               e^{-i{\bf k}\cdot {\bf r}} d^3{\bf r},  \label{eq:cor2sp}
\eea
where $<...>_x$ denotes an average over ${\bf x}$.
A Gaussian random variable satisfies
\be
        < B^2({\bf x})B^2({\bf x}+{\bf r}) >
     = < B^2({\bf x})><B^2({\bf x}+{\bf r}) > 
      +2<B({\bf x}) B({\bf x}+{\bf r})>^2,
\ee
where the first term on the right is a constant.
Therefore we can ignore the term in what follows.
Fourier transform of both sides results in
\bea
   \mbox{LHS} & = & E_{B^2, 3D}({\bf k}), \\
   \mbox{RHS} & = & 2 \int <B({\bf x})B({\bf x}+{\bf r})>^2 
                  e^{-i{\bf k}\cdot {\bf r}} d^3{\bf r} \\
          & = & 2\int~d^3{\bf r} ~{\cal K}_{B}({\bf r})~{\cal K}_{B}({\bf r}) 
                  e^{-i{\bf k}\cdot {\bf r}}  \\
          & = & 2\int~d^3{\bf r} \int~d^3{\bf p} \int~d^3{\bf q}
                  ~E_{B, 3D}({\bf p})E_{B, 3D}({\bf q})
                   e^{i({\bf p}+{\bf q}-{\bf k})\cdot {\bf r}}  \\
          & = & 2 \int~d^3{\bf p} \int~d^3{\bf q}
                  ~E_{B, 3D}({\bf p})E_{B, 3D}({\bf q})
                   \delta( {\bf p}+{\bf q}-{\bf k})  \\
          & = & 2\int~d^3{\bf p} ~E_{B, 3D}({\bf p})~E_{B, 3D}({\bf k}-{\bf p}),
\eea
where $\delta(k)$ is the Dirak $\delta$-function.
Therefore we have
\be
    E_{B^2, 3D}({\bf k}) \approx E_{B^2, 3D}({\bf 0})
    \approx 2\int~d^3{\bf k} | E_{B, 3D}({\bf k}) |^2
    \approx \mbox{constant}
\ee
for $k \ll k_0$.


\begin{thebibliography}{8.}



\bibitem{} Armstrong, J., Rickett, B., \& Spangler, S. 1995, ApJ, 443, 209

\bibitem{} Baccigalupi, C., Burigana, C., Perrotta, F., De Zotti, G.,
           La Porta, L., Maino, D., Maris, M., \& Paladini, R. 2001,
           A\&A, 372, 8
\bibitem{} Beuermann, K., Kanbach, G., \& Berkhuijen, E. 1985, A\&A, 153, 17
\bibitem{} Beresnyak, A. \& Lazarian, A. 2006, ApJL, 640, 175
\bibitem{} Boldyrev, S. 2002, ApJ, 569, 841
\bibitem{} Boldyrev, S. 2006, Phys. Rev. Lett., 96, 115002
\bibitem{} Bouchet, F. \& Gispert, R. 1999, New Astronomy, 4, 443
\bibitem{} Bouchet, F., Gispert, R., \& Pouget, J.-L. 1996, AIP Conf. Proc.,
           No. 348, 255

\bibitem{} Chepurnov, A. V. 2002, astro-ph/0206407 
           (Astron.Astrophys.Trans. 17 (1999) 281)


\bibitem{} Cho, J. \& Lazarian, A. 2002, ApJ, 575, 63 (CL02)

\bibitem{} de Oliveira-Costa, A., Tegmark, M., O'dell, C., 
           Keating, B., Timbie, P., Efstathiou, G., \& Smoot, G.
           2003, Phys. Rev. D, 68, 083003




\bibitem{} Dolginov, A.~Z., Gnedin, Iu.~N., \& Silantev, N.~A. 1996,
             {\it Propagation and Polarization of Radiation in Cosmic Media},
             (Gordon \& Breech)
\bibitem{} Draine, B. 1985, ApJS, 57, 587
\bibitem{} Draine, B. \& Lee, H. 1984, ApJ, 285, 89
\bibitem{} Draine, B. \& Flatau P. 1994, Opt. Soc. Am. A, 11, 1491
\bibitem{} Draine, B. \& Flatau P. 2008, arXiv:0809.0337
\bibitem{} Dunkley et al. 2008, arXive:0811.3915

\bibitem{} Finkbeiner, D. P., Davis, M., \& Schlegel, D. J. 1999, ApJ, 524, 867


\bibitem{} Fosalba, P., Lazarian, A., Prunet, S., \& Tauber, J.A. 2002
             ApJ, 564, 762 (FLPT)

\bibitem{} Getmantsev, G. 1959, Soviet Astronomy, 3, 415

\bibitem{} Giardino, G., Banday, A. J., Bennett, K., Fosalba, P., 
           Gorski, K. M., O’Mullane, W., Tauber, J., \& Vuerli, C. 
           2001a, in {\it Mining the Sky}, 
           ed. A. J. Banday, S. Zaroubi, \& M. L. Bartelmann 
           (Heidelberg: Springer), 458
\bibitem{} Giardino, G., Banday, A. J., Fosalba, P., Gorski, K. M., 
           Jonas, J. L., O’Mullane, W., \& Tauber, J. 2001b, 
           A\&A, 371, 708

\bibitem{} Giardino, G., Banday, A.J., G\'{o}rski, K.M., Bennett, K.,
             Jonas, J.L., \& Tauber, J. 2002, A\&A, 387, 82

\bibitem{} G\'{o}rski, K., Hivon, E., Banday, A., et al. 2005, ApJ, 622, 759

\bibitem{} Gustafsson, M., Brandenburg, A., Lemaire, J., \& Field, D.
           2006, A\&A, 454, 815

\bibitem{} Haslam, C., Salter, C., Stoffel, H., \& Wilson, W. 1982, A\&AS, 47, 1


\bibitem{} Heiles, C. 2000, AJ, 119, 923
\bibitem{} Hildebrand, R., Davidson, J., Dotson, J., Dowell, C.,
           Novak, G., \& Vaillancourt, J. 2000, PASP, 112, 1215
\bibitem{} Hildebrand, R., Dotson, J., Dowell, C., Schleuning, D., \&
           Vaillancourt, J. 1999, ApJ, 516, 834


\bibitem{} Jonas, J. L., Baart, E. E., \& Nicolson, G. D. 1998, 
            MNRAS, 297, 977
\bibitem{} Kowal, G., Lazarian, A., \& Beresnyak, A. 2007, ApJ, 658, 423

\bibitem{} Lazarian, A. 1992, Astron.~\& Astrophys.~Trans., 3, 33
\bibitem{} Lazarian, A. 1995a, Ph.~D.~Thesis (Univ. of Cambridge, UK)
\bibitem{} Lazarian, A. 1995b, A\&A, 293, 507
\bibitem{} Lazarian, A. 2007, J. Quant. Spectrosc. Radiat. Trans., 106, 225
\bibitem{} Lazarian, A. 2009, Space Science Rev., accepted (arXiv:0811.0839)
\bibitem{} Lazarian, A. \& Pogosyan, D. 2000, ApJ, 537, 720
\bibitem{} Lazarian, A. \& Prunet, S. 2001, in
             {\it Astrophysical Polarized Backgrounds}, AIP Conf.~Proc.~609, 
             ed.~S.~Cecchini et al.
             (Melville: AIP), 32

\bibitem{} Lazarian, A. \& Shutenkov, V.~P. 1990, PAZh, 16,
690 (translated Sov.~Astron.~Lett., 16, 297)


\bibitem{} La Porta, L., Burigana, C., Reich, W., Reich, P. 2008,
           A\&A, 479, 641
\bibitem{} La Porta, L., Burigana, C., Reich, W., Reich, P. 2006,
           A\&A, 455, 9

\bibitem{} Lee, H. \& Draine, B. 1985, ApJ, 290, 211

\bibitem{} Masi, S. et al. 2001, ApJL, 553, 93

\bibitem{} Martin, P. 1974, ApJ, 187, 461
\bibitem{} Mishchenko, M. I. 2000, Appl. Opt., 39, 1026
\bibitem{} Miville-Deschenes, M.-A., Ysard, N.,  Lavabre, A., Ponthieu, N., 
           Macias-Perez, J., Aumont, J., \& Bernard, J. 2008, A\&A, accepted 
           (arXiv:0802.3345)
\bibitem{} M\"uller, W.-C. \& Biskamp, D. 2000,
                Phys.~Rev.~Lett. 84(3), 475

\bibitem{} Page, L., Hinshaw, G., Komatsu, E., et al. 2007, ApJS, 170, 335

\bibitem{} Padoan, P., Boldyrev, S., Langer, W., \& Nordlund, A. 2003
           ApJ, 583, 308

\bibitem{} Ponthieu, N. et al. 2005, A\&A, 444, 327

\bibitem{} Prunet, S., Sethi, S., Bouchet, F., \& Miville-Deschenes, M.-A.
           A\&A, 339, 187
\bibitem{} Prunet, S. \& Lazarian, A. 1999,
           in {\it Microwave Foregrounds}, ASP Conf. Ser. 181, 
           ed.~A.~de~Oliveira-Costa and M.~Tegmark, p113

\bibitem{} Reich, P., \& Reich, W. 1986, A\&AS, 63, 205

\bibitem{} Schlegel, D., Finkbeiner, D., \& Davis, M. 1998, ApJ, 500, 525
\bibitem{}  She, Z.-S., Leveque, E. 1994, 
                Phys.~Rev.~Lett.,
                72(3), 336 
\bibitem{} Smoot, G.~F. 1999, in {\it Microwave Foregrounds}, 
             ASP Conf.~Ser.~181, 
             ed. A. de Oliveira-Costa \& M. Tegmark (San Francisco: ASP), 61


\bibitem{} Sun, X. H., Reich, W., Waelkens, A., \& Ensslin, T. A. 2008, A\&A, 477, 573

\bibitem{} Szapudi, I., Prunet, S., Pogosyan, D., Szalay, A. S., \&
             Bond, J. R. 2001, ApJ, 548, L115

\bibitem{} Tegmark, M. \& Efstathiou, G. 1996, MNRAS, 281, 1297
\bibitem{} Tegmark, M., Eisenstein, D.~J., Hu, W., de Oliveira-Costa, A.
             2000, ApJ, 530, 133

\bibitem{} Tucci, M., Carretti, E., Cecchini, S., Nicastro, L., Fabbri, R.,
           Gaensler, B. M., Dickey, J. M., McClure-Griffiths, N. M. 2002,
           ApJ, 579, 607

\bibitem{} Waelkens, A., Jaffe, T., Reinecke, M., Kitaura, F., \& Ensslin, T.
           2008, A\&A, submitted (arXive:0807.2262)

\bibitem{} Weingartner, J. \& Draine, B. 2001, ApJ, 548, 296
\bibitem{} Whittet, D., Hough, J., Lazarian, A., \& Hoang, T. 2008,
           ApJ, 674, 304

\end{thebibliography}
\end{document}